\documentclass[iop]{emulateapj}
\usepackage{textcomp}
\usepackage{epsfig, epstopdf}
\usepackage{amssymb,amsmath}
\usepackage{longtable,lscape}
\usepackage[]{natbib}
\newcommand{\kms}{\hbox{km~s$^{-1}$}}
\newcommand{\cmsq}{\hbox{cm$^{-2}$}}

\newcommand{\flux}{\hbox{erg~cm$^{-2}$~s$^{-1}$}}
\newcommand{\lumin}{\hbox{erg~s$^{-1}$}}

\newcommand{\aox}{\hbox{$\alpha_{\rm ox}$}}
\newcommand{\nh}{\hbox{${N}_{\rm H}$}}

\newcommand{\RF}{\hbox{rest-frame}}

\newcommand{\XR}{\hbox{X-ray}}
\newcommand{\XRs}{\hbox{X-rays}}
\newcommand{\PL}{\hbox{power-law}}
\newcommand{\PLs}{\hbox{power-laws}}
\newcommand{\asca}{{\emph{ASCA}}}
\newcommand{\chandra}{\emph{Chandra}}

\newcommand{\sax}{{\emph{BeppoSAX}}}
\newcommand{\rosat}{{\emph{ROSAT}}}
\newcommand{\einstein}{{\emph{Einstein}}}

\newcommand{\xmm}{\hbox{\emph{XMM-Newton}}}
\newcommand{\nustar}{\hbox{\emph{NuSTAR}}}
\newcommand{\astroh}{\hbox{\emph{ASTRO-H}}}
\newcommand{\athena}{\hbox{\emph{Athena}}}

\newcommand{\SB}{\hbox{0.5--2~keV}}
\newcommand{\HB}{\hbox{2--8~keV}}
\newcommand{\FB}{\hbox{0.5--8~keV}}

\newcommand{\xspec}{{\sc xspec}}

\newcommand{\irasa}{IRAS~07598$+$6508}
\newcommand{\irasb}{IRAS~14026$+$4341}
\newcommand{\pga}{PG~0946$+$301}
\newcommand{\pgb}{PG~1001$+$054}
\newcommand{\pgc}{PG~1004$+$130}

\newcommand{\pge}{PG~1700$+$518}
\newcommand{\pgf}{PG~2112$+$059}
\newcommand{\Qa}{Q~1246$-$057}
\newcommand{\SBS}{SBS~1542+541}
\newcommand{\mrk}{Mrk~231}
\newcommand{\cso}{CSO~755}

\begin{document}

\title{THE LONG-TERM X-RAY VARIABILITY OF BROAD ABSORPTION LINE QUASARS}

\author{C.~Saez\altaffilmark{1,2,3}, W.~N.~Brandt\altaffilmark{1,2},  S.~C.~Gallagher\altaffilmark{4}, F.~E. Bauer\altaffilmark{3,5}, and G.~P.~Garmire\altaffilmark{1}}

\altaffiltext{1}{Department of Astronomy \& Astrophysics, The Pennsylvania State University,
University Park, PA 16802, USA }

\altaffiltext{2}{Institute for Gravitation and the Cosmos, The Pennsylvania State University, University Park, PA 16802, USA}

\altaffiltext{3}{Pontificia Universidad Cat\'olica de Chile, Departamento de Astronom\'ia y Astrof\'isica, Casilla 306, Santiago 22, Chile }

\altaffiltext{4}{Department of Physics and Astronomy, The University of Western Ontario, London, ON N6A 3K7, Canada}

\altaffiltext{5}{Space Science Institute, 4750 Walnut Street, Suite 205, Boulder, CO 80301, USA}

\keywords{cosmology: observations
--- X-rays: galaxies --- galaxies: active --- quasars: absorption lines}

\begin{abstract}

We analyze the long-term (rest-frame \hbox{3--30~yr}) \XR\ variability 
of eleven broad absorption line (BAL) quasars, mainly to constrain
the variation properties of the	\XR\ absorbing shielding gas that 
is thought to play a critical role in BAL wind launching. Our 
BAL quasar sample has coverage with multiple \hbox{X-ray} 
observatories including \chandra, \xmm, \sax, \asca, \rosat, and 
\einstein; \hbox{3--11} observations are available for each source. 
For seven of the eleven sources we have obtained and analyzed new 
\chandra\ observations suitable for searching for any strong X-ray
variability. We find highly significant \XR\ variability in 
three sources (\pgb, \pgc, and \pgf). The maximum observed
amplitude of the \hbox{2--8~keV} variability is a factor of 
$3.8\pm1.3$, $1.5\pm0.2$, and $9.9\pm2.3$  for \pgb, \pgc, 
and \pgf, respectively, and these sources show detectable
variability on rest-frame timescales down to 5.8, 1.4, and 0.5~yr.  
For \pgc\ and \pgf\ we also find significant \XR\ spectral 
variability associated with the flux variability. Considering 
our sample as a whole, we do not find that BAL quasars 
exhibit exceptional long-term \XR\ variability when compared to the 
quasar population in general. We do not find evidence for common
strong changes in the shielding gas owing to physical rearrangement
or accretion-disk rotation, although some changes are found; this 
has implications for modeling observed ultraviolet BAL variability. 
Finally, we report for the first time an \XR\ detection of the 
highly polarized and well-studied BAL quasar IRAS~14026+4341 in 
its new \chandra\ observation. 

\end{abstract}

\section{INTRODUCTION}\label{S:intr}

Fast outflows are common in active galactic nuclei (AGNs) over a range 
of more than 10,000 in AGN luminosity \citep[e.g.,][]{2002ApJ...569..641L, 2008ApJ...672..102G}. 
These  outflows are a significant component of the nuclear environment, and 
their ubiquity suggests that mass ejection 
in a wind is fundamentally linked to mass accretion
onto a supermassive black hole (SMBH). Indeed, outflows may be key in allowing 
the accreting matter to expel angular momentum. Observationally, while 
these outflows are most commonly studied via the absorption features 
they imprint upon ultraviolet (UV) and \hbox{X-ray} spectra, they are 
also thought to shape emission-line and continuum properties 
\citep[e.g.,][]{1997ApJ...474...91M, 2011AJ....141..167R}. Studies 
of the outflows from luminous AGNs, such as those seen in Broad 
Absorption Line (BAL) quasars, have taken on further importance in 
recent years as they are one of the possible agents of feedback in 
typical massive galaxies  \citep[e.g.,][]{2005Natur.433..604D}. 
Outflows from luminous AGNs may expel enough gas from the 
SMBH/galaxy system to quench both SMBH growth and star 
formation, perhaps leading to the observed relations between SMBH 
and bulge properties  \citep[e.g.,][]{2009ApJ...698..198G}. 

One frequently used and generally well-supported model proposes that 
the UV absorption lines of BAL quasars originate in a line-driven 
wind that is launched from the accretion disk at 
$\approx 10^{16-17}$~cm from the SMBH \citep[e.g.,][]{1995ApJ...451..498M, 2000ApJ...543..686P}. 
In this model, the weak \hbox{X-ray} emission seen from BAL quasars
(e.g., Gallagher et~al. 2006; Gibson et~al. 2009)
is due to absorption in stalled, highly ionized 
``shielding gas'' at smaller radii that protects the wind from nuclear 
extreme UV and soft \hbox{X-ray} radiation; without such an absorbing layer, 
these energetic photons would so highly ionize the potential wind material that 
it could not be effectively accelerated via line pressure. 
Supporting observational evidence for the physical importance of 
the shielding gas comes in the form of correlations between the 
\hbox{X-ray} weakness of BAL quasars and the strength and velocity 
of their UV C~{\sc iv} BALs \citep[e.g.,][]{2006ApJ...644..709G, 2009ApJ...692..758G, 
2010ApJ...724..762W}. 

The nature of the shielding gas in BAL quasars remains poorly constrained 
observationally. Heavy \hbox{X-ray} absorption is directly seen
in many of the BAL quasars with \hbox{X-ray} spectra of reasonable quality
\citep[e.g.,][]{2001ApJ...546..795G, 2002ApJ...567...37G, 2009ApJ...690.1006F}. Measured 
absorption column densities span a broad range of typically
\hbox{$N_{\rm H}\approx 10^{22}$--$10^{24}$~cm$^{-2}$}, although 
most individual $N_{\rm H}$ measurements have significant uncertainties 
owing to limited photon statistics and absorption complexity
(e.g., ionization and partial-covering effects). The kinematics
of the \hbox{X-ray} absorber in BAL quasars is also poorly constrained. 
The \XR\ absorber is sometimes envisioned to be a stalled ``sacrificial wind'' that
protects the UV-absorbing wind launched at larger radii
\citep[e.g.,][]{2000ApJ...543..686P}, although in one BAL quasar apparent evidence 
for a rapidly outflowing \hbox{X-ray} absorber has been found 
\citep[e.g.,][]{2009ApJ...706..644C, 2011ApJ...737...91S}.  Magnetocentrifugal effects, as well as radiative ones, may play an important role in setting the kinematics of the \XR\ absorbing shielding gas \citep[e.g.,][]{2005ApJ...631..689E}.

Remarkable \hbox{X-ray} absorption variability has been seen in a few 
BAL quasars \citep[e.g.,][]{2004ApJ...603..425G, 2005A&A...433..455S, 2007A&A...474..431S, 2006ApJ...652..163M, 2009ApJ...706..644C},
typically on multi-year timescales, indicating that the \hbox{X-ray} 
absorber is indeed a dynamical structure. This variability could 
be due simply to accretion-disk rotation moving an 
azimuthally asymmetric \hbox{X-ray} absorber through our 
line of sight; at the expected radius of the
shielding gas ($\lesssim 10^{15-16}$~cm), substantial fractional 
disk rotation is expected on multi-year timescales.  
The results on \hbox{X-ray} absorption variability in BAL quasars 
are broadly reminiscent of those found for lower luminosity 
Seyfert~2 galaxies, where \hbox{X-ray} absorbing column densities are 
found to vary commonly \citep[e.g.,][]{2002ApJ...571..234R, 2009MNRAS.393L...1R}.
In addition to variations in X-ray absorber column density, some BAL
quasars with multiple observations have also shown notable continuum luminosity
variability. For example,  the normalization of the power-law continuum in
PG~2112+059 decreased by a factor $\approx 3.5$ between 1999 and 2002
\citep{2004ApJ...603..425G}, and continuum variability of this magnitude
continued through 2007 \citep{2010A&A...512A..75S}.  

Current studies of the \hbox{X-ray} absorption variability of BAL quasars 
suffer from small sample sizes. Presently only a handful of BAL quasars 
have sensitive coverage in multiple \hbox{X-ray} observations, and thus the 
general frequency of BAL quasar \hbox{X-ray} variability remains poorly 
characterized. In order to improve this situation, we targeted seven
optically bright BAL quasars as part of \chandra\ Cycle~11 that already
had sensitive archival \hbox{X-ray} observations with \chandra, \xmm, or 
earlier missions. We obtained relatively short observations \hbox{(5--7~ks)} 
suitable for searching for any strong \hbox{X-ray} variability. Our main goal was 
to execute an economical multi-year \hbox{X-ray} variability search for as many 
objects as possible, so that the basic \hbox{X-ray} variability properties
of BAL quasars could be characterized as a class. Similar ongoing 
efforts are also now characterizing the multi-year UV absorption-line
variability of BAL quasars \citep[e.g.,][]{2008ApJ...675..985G, 2010ApJ...713..220G, 2011MNRAS.413..908C, 2012MNRAS.422.3249C, 2012arXiv1208.0836F}, and  \XR\ constraints upon shielding gas variations will inform these efforts. In addition to presenting our new \chandra\  Cycle~11 observations below, we also utilize the existing archival \hbox{X-ray}  observations of BAL quasars capable of constraining long-term \hbox{X-ray} 
variability; this maximizes our sample size.

The layout of this paper is as follows: in \S \ref{S:data} we describe the sample and  the \XR\ data reduction;  in \S \ref{S:span} we provide a description of the spectral analysis; in \S \ref{S:resu} we provide our variability results and discussion focusing on \hbox{long-term} \XR\ variability; and in \S \ref{S:conc} we summarize our main findings.  Throughout this paper, unless stated
otherwise,  we use cgs units, the errors
listed are at the 1$\sigma$ level, and we adopt a flat
$\Lambda$-dominated universe with $H_0=70~\kms$~Mpc$^{-1}$,
$\Omega_\Lambda=0.7$, and $\Omega_M=0.3$.

\begin{deluxetable*}{cccccccccccccccc}
\tablecolumns{13}
\tabletypesize{\scriptsize}
 \tablewidth{0pt}
\tablecaption{Optical/Radio Properties, Galactic Column Densities, and 
\XR\ Coverage of Surveyed BAL Quasars  \label{tab:opra}}
\tablehead
{
\colhead{\sc object name} &
\colhead{$z$} &
\colhead{$\alpha_{2000.0}$\tablenotemark{a}} &
\colhead{$\delta_{2000.0}$\tablenotemark{a}} &
\colhead{\nh$_{\rm Gal}$\tablenotemark{b}}  &
\colhead{AB$_{2500}$\tablenotemark{c}} &
\colhead{log~$R_L$\tablenotemark{d}} &
\colhead{BAL type\tablenotemark{e}} &
\colhead{$N_{\XR}$\tablenotemark{f}} &
\colhead{$\Delta t_{\rm rest}$\tablenotemark{f}} 
}
\startdata


\irasa  & 0.148  & 08 04 30.51 & $+$64 59 52.7 & 4.3 &  14.5 & $0.80$ & Lo  & 5 & 16.8  \\

\pga  &1.221 &  09 49 41.10 & $+$29 55 19.2   & 1.7  &  16.2 & $-1.71$ & Hi & 3 & 7.3   \\

\pgb  & 0.160 & 10 04 20.14  & $+$05 13 00.5  &  2.4  & 16.6 & $-0.12$ & Hi & 5 & 26.0  \\

\pgc  & 0.241 & 10 07 26.11 & $+$12 48 56.1  &  3.7 & 15.3 & $2.18$ & Hi & 3 & 19.9 \\

\Qa & 2.247  & 12 49 13.86 &  $-$05 59 19.1 & 2.1 & 16.3 & $<$$-0.14$ & Hi & 8 & 7.7  \\

\mrk & 0.042  & 12 56 14.23 & $+$56 52 25.2 &  1.3 & 15.3 & $1.41$  & Lo & 11 & 28.4   \\

\irasb  &  0.323 & 14 04 38.80 & $+$43 27 07.4   &  1.2 & 16.2 & $-0.27$ & Lo & 4 & 13.6  \\ 

\cso & 2.882  & 15 25 53.89   & $+$51 36 49.2  &  1.6  & 16.2 & $<$$-0.19$ & Hi   & 5 & 2.9 \\
\SBS & 2.370 & 15 43 59.45 &  $+$53 59 03.4 & 1.3 & 16.8 & $-0.85$  & Hi  & 4 & 2.5    \\
\pge & 0.292 & 17 01 24.80 & $+$51 49 20.0  & 2.6   & 15.3 & $0.38$ & Lo & 6 & 15.0  \\ 
\pgf &  0.466 & 21 14 52.57  & $+$06 07 42.5   &  6.6 & 15.0 & $-0.56$ & Hi & 9 & 10.9  \\

\enddata

\tablenotetext{a}{Optical positions in J2000.0 equatorial coordinates.}

\tablenotetext{b}{Galactic absorption column densities in units of $10^{20}$~\cmsq\
\citep[from][]{1990ARA&A..28..215D}, obtained using the optical coordinates of the
sources and the HEASARC \nh\ tool at http://heasarc.gsfc.nasa.gov/cgi-bin/Tools/w3nh/w3nh.pl}

\tablenotetext{c}{The monochromatic AB magnitude at rest-frame wavelength 2500 \AA.  These were computed from the  flux densities obtained from the closest optical magnitude found in the NASA Extragalactic Database (NED; http://ned.ipac.caltech.edu)  and  extrapolating a \PL\ with $\alpha=-0.5$ \citep[e.g.,][]{2001AJ....122..549V}.  Previous to the extrapolation the flux densities obtained through NED have been corrected for Galactic reddening.}






\tablenotetext{d}{\hbox{Logarithm} of the \hbox{radio-loudness} parameter ($R_L=f_{\rm 5GHz}/f_{\rm 4400\AA}$). The flux densities at \RF\ 4400 \AA\ are obtained  using the Galactic reddening corrected flux densities obtained from the closest optical magnitudes and extrapolating a \PL\ with $\alpha=-0.5$ \citep[e.g.,][]{2001AJ....122..549V}. The flux densities at \RF\ 5~GHz have been obtained using either the flux densities at 1.4~GHz or the flux densities at 5~GHz. When  both the 1.4~GHz and 5~GHz flux densities are available, we find the flux density at \RF\ 5~GHz through a \PL\ interpolation between the two values. For all the sources with the exception of \SBS\ the flux densities at 1.4~GHz were obtained from the NVSS \citep{1998AJ....115.1693C} or the FIRST survey \citep{1997ApJ...475..479W}. Additionally for PG sources the flux densities at 5~GHz were obtained from  \cite{1989AJ.....98.1195K}. For \SBS\ the flux densities at 1.4~GHz and 5~GHz were obtained from \cite{2006AJ....132.1307P}.}


\tablenotetext{e}{Hi = ultraviolet spectra show high-ionization BALs only; Lo = low-ionization (Mg {\sc ii}  and/or Al {\sc iii}) BALs present.}

\tablenotetext{f}{Number of \XR\ observations ($N_{\XR}$) with their \RF\ time span coverage in years ($\Delta t_{\rm rest}$).}

\end{deluxetable*}
  
\begin{deluxetable*}{cccccc}
\tablecolumns{7}
\tabletypesize{\scriptsize}
 \tablewidth{0pt}
\tablecaption{Log of \chandra, \sax, \rosat, and \einstein\ Observations \label{tab:chao}}
\tablehead
{
\colhead{} & \colhead{} & \colhead{{\sc exp. time}\tablenotemark{a}}  &  \colhead{} & \colhead{\sc rate} \\
\colhead{\sc object name} & \colhead{\sc obs. date} & \colhead{(ks)} & \colhead{\sc counts$^{\rm a,b}$ ~~}  & \colhead{(10$^{-3}$~s$^{-1}$)} & \colhead{Ref.\tablenotemark{c}}
}
\startdata
\\
\multicolumn{6}{c}{\small  New \chandra\ observations (\FB)} \\
 \\
\hline
\irasa  & 2010 Jun 18 & 6.7 & $32_{-6}^{+7}$ & $4.7_{-0.8}^{+1.0}$  &  1  \\
\pga & 2010 Jan 11 & 6.5 & $57_{-8}^{+9}$ & $8.8_{-1.2}^{+1.3}$  & 1  \\ 
\pgb  & 2010 Jan 11  &  4.9 & $13_{-4}^{+5}$ & $2.6_{-0.7}^{+0.9}$ & 1  \\
\mrk & 2010 Jul 11 & 4.8 & $247_{-18}^{+19}$ & $51.6_{-3.7}^{+3.9}$  & 1  \\
\irasb & 2010 Jul 28  & 6.7 & $6_{-2}^{+4}$ & $0.9_{-0.3}^{+0.5}$ & 1  \\ 

\cso  & 2010 Mar 21& 4.9  & $110_{-10}^{+12}$ & $22.4_{-2.1}^{+2.3}$ & 1  \\
\pge & 2010 Jun 26 & 6.7 & $14_{-4}^{+5}$ & $2.0_{-0.5}^{+0.7}$  &1  \\ 

\hline
\\
\multicolumn{6}{c}{\small Archival \chandra\ observations  (\FB)} \\
\\
\hline
\irasa & 2000 Mar 21 & 1.3 & $9_{-3}^{+4}$ & $6.7_{-2.2}^{+3.1}$  &  2  \\
\pgc  & 2005 Jan 5 & 41.1 & $1905_{-46}^{+47}$ & 46.4$\pm$1.1  & 3  \\
\Qa & 2000 Feb 8 & 5.5 & $42_{-6}^{+8}$ & $7.6_{-1.2}^{+1.4}$  & 2\\
\mrk   & 2000 Oct 19 & 39.3 & $2059_{-52}^{+53}$ & 52.5$\pm$1.3  &  4  \\
\mrk  & 2003 Feb 3 & 39.7 & $1947_{-47}^{+48}$ & 49.1$\pm$1.2  &  5  \\
\mrk  & 2003 Feb 11 & 38.6 & $1900_{-47}^{+48}$ & 49.2$\pm$1.2  & 5  \\ 
\mrk & 2003 Feb 20 & 36.0 & $1754_{-46}^{+47}$ & 48.7$\pm$1.3  & 5  \\
\SBS & 2000 Mar 22 & 4.5 & $79_{-9}^{+10}$ & $17.3_{-1.9}^{+2.2}$  & 2\\
\pgf  & 2002 Sep 1 & 56.9 & $836_{-29}^{+30}$ & 14.7$\pm$0.5  & 6  \\

\hline
\\
\multicolumn{6}{c}{\small Archival \sax\ observations (1.8$-$10~keV)} \\
\\
\hline
 \mrk & 2001 Dec 29 & 144.0 &  818$\pm$36 & 5.7$\pm$0.3 & 7\\
 \cso &  1999 Feb 02  &   35.2 & 53$\pm$18& 1.5$\pm$0.5 & 8 \\
\hline
\\
\multicolumn{6}{c}{\small Archival \rosat\ observations (\SB)} \\
\\
\hline

\irasa  & 1991 Mar 17  & 7.9 & $18_{-6}^{+7}$ & 2.2$\pm$0.8 & 9  \\
\pga  & 1993 Nov 14  & 11.8 & $<$7 & $<$0.6 & 1  \\ 
\pgb   & 1992 May 17  & 8.6 & $9_{-4}^{+5}$ & 1.1$\pm$0.5 & 10  \\ 

\Qa & 1991 Dec 21 & 12.8 & 20$\pm$8 & 1.6$\pm$0.6 & 1  \\ 
\Qa & 1992 Jul 17  & 2.8 & $<$4 & $<$1.5 & 11  \\ 
\Qa & 1993 Jan 5  & 9.1& $14_{-5}^{+6}$ & $1.6_{-0.5}^{+0.6}$ & 11  \\ 
\Qa & 1993 Jun 22  & 43.6 & $59_{-12}^{+13}$ & 1.4$\pm$0.3 & 11  \\ 
 
 \mrk & 1991 Jun 7  & 23.2 & $224_{-17}^{+18}$ & 9.7$\pm$0.8 & 12  \\
\irasb  & 1992 Jul 18  & 6.2 & $<$5 & $<$0.8 & 1 \\ 
 \SBS & 1993 Ago 14  & 5.5 & $25_{-6}^{+7}$ & $4.5_{-1.0}^{+1.2}$ & 11  \\
 \pge  & 1991 Feb 9  & 7.8 & $<$7 & $<$0.9 & 11  \\ 
 \pgf  & 1991 Nov 30  & 11.7 & $52_{-8}^{+9}$ & $4.5_{-0.7}^{+0.8}$ & 13  \\

 \hline
 \\
\multicolumn{6}{c}{\small Archival \einstein\ observations (0.16$-$3.5~keV)} \\
\\
\hline
\pgb & 1979 Dec 02 & 1.6   &  $<$15 & $<$9.4 & 14 \\
\pgc & 1979 May 21 &7.1 &  $<$37 & $<$4.9 & 14, 15\\
\Qa & 1979 Jul 20 & 3.5 & $<$24 & $<$7.0 & 14\\
 \mrk & 1980 Dec 23 & 2.1 &  $<$19 & $<$ 8.1 & 14\\

\enddata
\tablenotetext{a}{The counts correspond to the net photon counts (background subtracted). The exposure times and photon counts are obtained after screening the data. }


 
\tablenotetext{b}{For details about the aperture sizes used see \S \ref{S:data}. 
The errors on the  source counts were computed by propagating the asymmetric errors on the total and background counts using the approach of \cite{2004physics...6120B}. The total and background count errors were obtained from Tables 1 and 2 of \cite{1986ApJ...303..336G}. }
\tablenotetext{c}{{\sc References:} (1)~This work; (2)~\cite{2001ApJ...558..109G}; (3)~\cite{2006ApJ...652..163M}; (4)~\cite{2002ApJ...569..655G}; (5)~\cite{2005ApJ...633...71G};  (6)~\cite{2004ApJ...603..425G};  (7)~\cite{2004A&A...420...79B};  (8)~\cite{1999ApJ...525L..69B}; (9)~\cite{1997MNRAS.285..879L}; (10)~\cite{1997ApJ...477...93L}; (11)~\cite{1996ApJ...462..637G}; (12)~\cite{1996MNRAS.278.1049R};  (13)~\cite{1996A&A...309...81W}; (14)~\cite{1994ApJS...92...53W}; (15)~\cite{1984ApJ...280...91E}. }

\end{deluxetable*}

\begin{deluxetable*}{cccccccccccc}
\setlength{\tabcolsep}{0.06in}
\tablecolumns{9}
\centering
\tabletypesize{\scriptsize}
 \tablewidth{0pt}
\tablecaption{Log of \xmm\ Archival Observations \label{tab:xmmo}}
\tablehead
{
\colhead{} & \colhead{} & \multicolumn{3}{c}{\underline{{\sc exposure time\tablenotemark{a}}}} & \multicolumn{2}{c}{\underline{\hspace{20pt}MOS1\tablenotemark{b}\hspace{20pt}}} & \multicolumn{2}{c}{\underline{\hspace{20pt}MOS2\tablenotemark{b}\hspace{20pt}}} & \multicolumn{2}{c}{\underline{\hspace{20pt}pn\tablenotemark{b}\hspace{20pt}}}  \\
 \colhead{} & \colhead{} & \colhead{MOS1} & \colhead{MOS2} & \colhead{pn}  & \colhead{\sc counts} &
\colhead{\sc rate$\times$10$^3$} & \colhead{\sc counts} & \colhead{\sc rate$\times$10$^3$} & \colhead{\sc counts} & \colhead{\sc rate$\times$10$^3$}   & \\
\colhead{\sc object name} & \colhead{\sc obs. date} & \multicolumn{3}{c}{(ks)}  & \multicolumn{2}{c}{(0.5$-$10~keV)} & \multicolumn{2}{c}{(0.5$-$10~keV)} & \multicolumn{2}{c}{(0.3$-$10~keV)} & \colhead{\sc ref.\tablenotemark{c} }
}
\startdata
\irasa & 2001 Oct 24 & 18.6 &18.5 & 14.2  & $84_{-11}^{+12}$ & 4.5$\pm$0.6 & $70_{-10}^{+11}$ & $3.8_{-0.5}^{+0.6}$ & $268_{-19}^{+20}$ & $18.9_{-1.3}^{+1.4}$  &  1  \\

\pgb & 2003 May 04	 & 11.3 & 11.3 & 8.8 & $78_{-10}^{+11}$ & 6.9$\pm$0.9 & $62_{-9}^{+10}$ & $5.5_{-0.8}^{+0.9}$ & 322$\pm$20 & $36.4_{-2.2}^{+2.3}$  &  2  \\

\pgc &  2003 May 04  &  21.5 & 21.5 & 18.5 &  $567_{-26}^{+27}$ & 26.4$\pm$1.2 & $622_{-27}^{+28}$ & $28.9_{-1.2}^{+1.3}$ & $1777_{-46}^{+47}$ & $98.3_{-2.5}^{+2.6}$  & 3\\

\Qa & 2001 Jul 11 & 40.0 & 40.0 & 33.5 & $126_{-14}^{+15}$ & 3.2$\pm$0.4 & $127_{-14}^{+15}$ & 3.2$\pm$0.4 & 335$\pm$28 & 10.0$\pm$0.8  &  4 \\
\Qa & 2004 Jul 15 & 46.4 & 47.0 & 34.8 & 141$\pm$16 & $3.0_{-0.3}^{+0.4}$ & $123_{-15}^{+16}$ & 2.6$\pm$0.3 & 351$\pm$25 & 10.1$\pm$0.7  &  5 \\


\mrk & 2001 Jun 07 & 21.4  & 21.4 & 17.2  & $774_{-29}^{+30}$ & 36.1$\pm$1.4 & $739_{-29}^{+30}$ & 34.5$\pm$1.4 & $2165_{-50}^{+51}$ & 125.8$\pm$2.9  &  6 \\

\irasb &  2005 Nov 26  & 8.7  & 7.9 & 0.0 & $<$10 & $<$1.2 & 14$\pm$6 & $1.8_{-0.7}^{+0.8}$ & ... & ...  & 7 \\
\irasb & 2006 Jan 16  & 6.9  & 6.2 & 0.0    & 12$\pm$6 & $1.7_{-0.8}^{+0.9}$ &  $<$8 & $<$1.3 & ... & ...  & 7 \\

\cso  & 2001 Jul 30  & 2.2  & 2.2 & 0.0  &  $29_{-6}^{+7}$ & $13.1_{-2.7}^{+3.1}$ & $41_{-7}^{+8}$ & $18.7_{-3.1}^{+3.5}$ & ... & ...  &  8 \\
\cso  & 2001 Dec 08 &  30.1  & 30.3 & 24.6  &  $610_{-27}^{+28}$ & 20.3$\pm$0.9 & $578_{-26}^{+27}$ & 19.1$\pm$0.9 & $1820_{-47}^{+48}$ & 73.8$\pm$1.9  & 8 \\
\cso  & 2001 Dec 13  & 4.5  & 3.7 & 2.0 & 74$\pm$10 & $16.3_{-2.1}^{+2.3}$ & $57_{-8}^{+9}$ & $15.2_{-2.2}^{+2.4}$ & $120_{-13}^{+14}$ & $61.0_{-6.7}^{+7.0}$  & 8 \\

\SBS & 2002 Feb 03 & 8.4 & 8.5 & 2.4 & $99_{-11}^{+12}$ & $11.7_{-1.3}^{+1.4}$ & $74_{-10}^{+11}$ & $8.7_{-1.2}^{+1.3}$ & $98_{-11}^{+12}$ & $40.4_{-4.6}^{+5.0}$  &  4 \\
\SBS & 2002 Feb 06 & 23.8 & 23.6 & 12.7 & $261_{-18}^{+19}$ & 11.0$\pm$0.8 & $268_{-18}^{+19}$ & 11.3$\pm$0.8 & $489_{-24}^{+25}$ & 38.5$\pm$1.9  &   4 \\


\pge & 2009 Dec 19  & 7.0  & 6.9 & 4.5 &  $<$7 & $<$0.9 & $<$8 & $<$1.2 & 30$\pm$9 & $6.7_{-1.9}^{+2.0}$  &  9 \\
\pge  & 2009 Dec 31  & 14.1  & 12.8 & 7.9 &   $<$12 & $<$0.9 & 14$\pm$7 & $1.1_{-0.5}^{+0.6}$ & $53_{-11}^{+12}$ & $6.7_{-1.4}^{+1.5}$  & 9 \\
\pge & 2010 Jan 02  & 19.3  &18.8 &15.4 &  $32_{-8}^{+9}$ & $1.7_{-0.4}^{+0.5}$ & 20$\pm$8 & 1.1$\pm$0.4 & 87$\pm$14 & 5.6$\pm$0.9  & 9 \\

\pgf & 2003 May 14  & 9.3  & 9.7 & 6.9 &  $340_{-20}^{+21}$ & $36.4_{-2.1}^{+2.2}$ & $343_{-20}^{+21}$ & 35.3$\pm$2.1 & $900_{-33}^{+34}$ & $130.0_{-4.8}^{+4.9}$  & 10 \\
\pgf  & 2005 Nov 20 & 72.9 & 73.1 & 64.8  & 245$\pm$19 & 3.4$\pm$0.3 & $235_{-18}^{+19}$ & $3.2_{-0.2}^{+0.3}$ & $785_{-33}^{+34}$ & 12.1$\pm$0.5  &  10 \\
\pgf  & 2007 May 03  & 23.7 & 23.2 & 19.3  & $132_{-13}^{+14}$ & 5.6$\pm$0.6 & 124$\pm$13 & $5.3_{-0.5}^{+0.6}$ & 336$\pm$23 & 17.4$\pm$1.2  &  10 \\
\pgf & 2007 May 19 & 59.3 & 74.2 & 43.9  & $293_{-20}^{+21}$ & $4.9_{-0.3}^{+0.4}$ & $372_{-22}^{+23}$ & 5.0$\pm$0.3 & $767_{-33}^{+34}$ & 17.5$\pm$0.8  &  10 \\
\pgf & 2007 May 21  & 85.7 & 87.5 & 66.3   & 494$\pm$26 & 5.8$\pm$0.3 & $456_{-24}^{+25}$ & 5.2$\pm$0.3 & $1289_{-41}^{+42}$ & 19.4$\pm$0.6  & 10 \\
\pgf & 2007 Nov 05   & 50.2 & 49.0 & 43.0  & 190$\pm$17 & 3.8$\pm$0.3 & 200$\pm$17 & $4.1_{-0.3}^{+0.4}$ & $590_{-29}^{+30}$ & 13.7$\pm$0.7  & 10 \\


\enddata



\tablenotetext{a}{The exposure times and photon counts are obtained after screening the data for flaring.} 



\tablenotetext{b}{For details about the source and background extraction regions see \S \ref{S:data}. The errors on the  source counts were computed by propagating the asymmetric errors on the total and background  counts using the approach of \cite{2004physics...6120B}. The total and background count errors were obtained from Tables 1 and 2 of \cite{1986ApJ...303..336G}. }

\tablenotetext{c}{{\sc references:} (1)~\cite{2004AJ....127..758I};  (2)~\cite{2005A&A...433..455S}; (3)~\cite{2006ApJ...652..163M}; (4)~\cite{2003AJ....126.1159G}; (5)~\cite{2011MNRAS.416.2792P}; (6)~\cite{2004A&A...420...79B}; (7)~\cite{2007A&A...471..775R}; (8)~\cite{2005AJ....130.2522S}; (9)~\cite{2011MNRAS.415.2600B}; (10)~\cite{2010A&A...512A..75S}.}

\end{deluxetable*}

\section{OBSERVATIONS AND DATA REDUCTION} \label{S:data}

Our sample consists of eleven BAL quasars that have been observed in the past by sensitive X-ray missions including \chandra, \xmm, \sax, \asca, \rosat, and \einstein. Seven of  these BAL quasars have been re-observed with new short-exposure (5--7~ks) \chandra\ observations that are analyzed in this paper. In Table~\ref{tab:opra} we  present the Galactic column densities, \hbox{radio-loudness} parameters ($R_L$),\footnote{The \hbox{radio-loudness} parameter is defined as the ratio between the flux density  at 5~GHz and the flux density  at 4400 \AA\  in the rest frame of the source \citep{1989AJ.....98.1195K}.} and some basic optical and radio properties for our sample. The redshifts of our sample quasars range from 0.04$-$2.88. The only sources that are not \hbox{radio-quiet} quasars (RQQs; $R_L \leq 10$) are \pgc\ and \mrk, \hbox{radio-loud} quasars (RLQs) with  $R_L > 10$.   Notice, however, that in the case of \mrk\ there is likely substantial reddening \citep[$A_V\sim2$; e.g.,][]{1977MNRAS.178..451B}; as a consequence the intrinsic 4400~\AA\ flux density would be approximately an order of magnitude higher.  Based on this argument  \mrk\  would not be considered a \hbox{bona-fide} RLQ. 

\begin{figure*}
   \includegraphics[width=14cm]{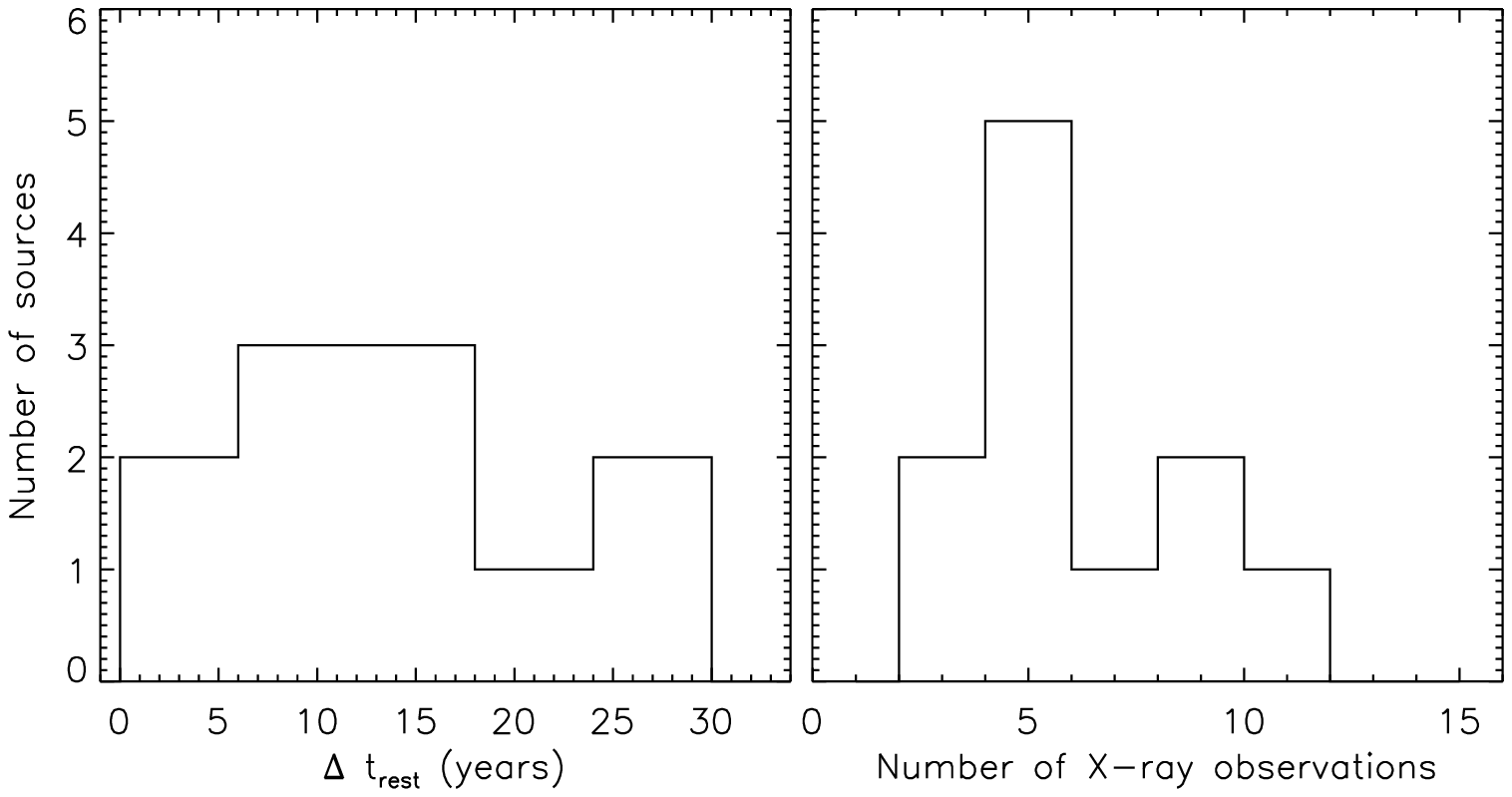}
        \centering
       \caption{Number of sources versus \RF\ time span of \XR\ coverage (left panel) and number of \XR\ observations (right panel).}
     \label{fig:dist}
     \end{figure*}

 In Table~\ref{tab:opra} we also present the number of \XR\ observations of each source and the \RF\ time span coverage.   The distributions of the number of X-ray observations with the respective \RF\ time span coverage are presented in Figure~\ref{fig:dist}. There are 3$-$11 \XR\ observations  with rest-frame time span ranging from 3$-$30 yr per source.
The observation logs of the sources analyzed in this work, which include observation dates, observed count rates  with 1$\sigma$ error bars, and exposure times, are presented in Table~\ref{tab:chao} for the \chandra, \sax, \rosat, and \einstein\ observations; Table~\ref{tab:xmmo} for the \xmm\ observations; and Table~\ref{tab:asca} for the \asca\ observations. The errors on the  source counts were computed by propagating the asymmetric errors on the total and background  counts using the approach described in \S5 of \cite{2004physics...6120B}. The total and background count errors were obtained from Tables 1 and 2 of \cite{1986ApJ...303..336G}.

\subsection{\chandra\ observations}
The \chandra\ observations (new and archival) were analyzed using the standard software CIAO version 4.2 provided by the \chandra\ \XR\ Center (CXC).  Standard CXC threads were employed to screen the data for status, grade (\emph{ASCA} grade 0, 2, 3, 4, and 6 events), and time intervals of acceptable aspect solution and background levels. For most of the observations the source and background spectra  were extracted (using the CIAO task {\sc dmextract}) from a circular region with an aperture radius of 4\arcsec\  \footnote{The 1.5 keV encircled energy at 4\arcsec\  from the center of an \hbox{on-axis} point  source is $\sim 99\%$ for ACIS-S .}  and an annular source-free region with an inner radius of 6\arcsec\ and an outer radius of  24\arcsec,  respectively.  
For the cases of PG 1004+130 and Mrk 231 the source, annulus inner, and annulus outer radii are (16\arcsec, 20\arcsec, 30\arcsec) and (24\arcsec, 30\arcsec, 50\arcsec), respectively.
For these sources we have selected larger extraction regions given that they possess extended \XR\ emission. The extended emission of \pgc\ and \mrk\ has been studied by \cite{2006ApJ...652..163M} and \cite{2002ApJ...569..655G}, respectively. We selected the radius of extraction for these sources using a fixed background region (annular region of inner radius 40\arcsec\ and outer radius 60\arcsec) and estimating the total counts as a function of aperture radius. The aperture radius selected  corresponds to that where the total counts reach an approximately constant value.  The use of larger source regions for the \chandra\ observations of \pgc\ and \mrk\ is necessary in order to make fair flux comparisons with the other X-ray missions that have more limited angular resolution.  The numbers of counts  in the source regions at energies of 0.5--8~keV  are presented in Table~\ref{tab:chao}.  The response files used to analyze the source spectra were created  with the CIAO tasks {\sc mkacisrmf} for the redistribution matrix files (RMFs)  and {\sc mkarf} for  the ancillary response files (ARFs). 

\subsection{\xmm\ observations}
The \xmm\ data were analyzed with the Science Analysis Software (SAS) version 10.0.0 provided by the \xmm\ Science Operations Centre (SOC).  The events files were obtained by including the events with ${\rm flag }=0$, ${\rm pattern} \leq 12$ (${\rm pattern} \leq 4$), and $200 \leq {\rm PI} \leq 12,000$ ($150 \leq {\rm PI} \leq15,000$) for the MOS (pn) detectors. The events files were also temporally  filtered in order to remove periods of flaring activity. To filter the events files for periods of flaring activity, we created light curves using single events with energies greater than 10~keV.  Using these light curves we find and remove from the events files the time periods where the count rate is above 0.5~count~s$^{-1}$ for the MOS cameras and 1~count~s$^{-1}$ for the pn camera.  
For the three EPIC detectors (MOS1,  MOS2, and pn), the source counts for each quasar were extracted from a circular region with an aperture radius of either 18\arcsec\ or 30\arcsec.\footnote{The 1.5 keV encircled energy at 18\arcsec\  (30\arcsec) from the center of an \hbox{on-axis} point  source is $\approx 75\%$ ($\approx 85\%$)  for the EPIC cameras.   The aperture corrections for the EPIC cameras are performed through the use of the {\sc arfgen} routine.} 
 The default aperture radius of the source extraction region was 30\arcsec; however, when the background counts were $>20\%$ of the total counts inside this region an aperture radius of 18\arcsec\ was used.
Under this criterion most observations were reduced using 18\arcsec\ apertures. The cases that were reduced with 30\arcsec\ apertures were the 2001 December 8th observation of \cso,  the \mrk\ observation, the \pgb\ observation, and the 2003 May 14 observation of \pgf.
The background spectra for most of the sources were extracted from annular source-free regions with 50$^{\prime\prime}$ inner radii and 100$^{\prime\prime}$  outer radii.  In cases where this region covered other X-ray sources or detector gaps, the background was extracted from a \hbox{source-free} circular region with radius 80\arcsec\ in the vicinity of the source. The numbers of counts  in the source regions for the \xmm\ cameras are presented in Table~\ref{tab:xmmo}.  These counts are selected at energies of 0.5--10~keV  for the MOS cameras and 0.3--10~keV for the pn camera. The response files used to analyze the source spectra were created with the SAS tasks {\sc rmfgen} for the RMFs and {\sc arfgen} for the ARFs. 

\subsection{\rosat\ observations}
The \rosat\ PSPC observations were analyzed following the \rosat\ {\sc xselect} guide.\footnote{http://heasarc.nasa.gov/docs/rosat/ros\_xselect\_guide} Using the prescription given we applied all the standard corrections (gain variations, gain saturation, analog-to-digital nonlinearity, and dead time) to obtain cleaned events which were used for the subtraction of the source and background spectra. The source spectra were extracted from  circular regions with aperture radii of 50\arcsec. The background spectra for most of the sources were extracted from annular source-free regions with 100\arcsec\ inner radii  and 200\arcsec\  outer radii; however, in cases where these regions were covering other X-ray sources, the background spectra were extracted from \hbox{source-free} circular regions with radii of 150\arcsec\ in the vicinity of the source.  The numbers of counts  in the source regions at energies of 0.5--2~keV  are presented in Table~\ref{tab:chao}.  The RMFs used for spectral fitting were obtained from the public calibration database.\footnote{ftp://legacy.gsfc.nasa.gov} The ARFs were generated with the {\sc pcarf} routine; this routine does not include aperture corrections, and therefore the \rosat\ fluxes were \hbox{aperture-corrected} using the integrated \rosat\ PSPC PSF. 

\begin{deluxetable*}{ccccccccc}
\tablecolumns{9}
\tabletypesize{\scriptsize}
 \tablewidth{0pt}
\tablecaption{Log of \asca\ Archival Observations \label{tab:asca}}
\tablehead
{
\colhead{} & \colhead{} & \multicolumn{2}{c}{\underline{{\sc exp. time(ks)\tablenotemark{a}}}} &  \multicolumn{2}{c}{\underline{{\sc \hspace{5pt} count rate($\times 10^3$)  \hspace{5pt}}}} \\
\colhead{\sc object name} &
\colhead{\sc obs. date} &
\colhead{SIS0} &
\colhead{SIS1} &
\colhead{SIS0} &
\colhead{SIS1} &
\colhead{\sc ref.$\rm ^b$} 
}
\startdata

\irasa &  1996 Oct 29 & 37.2 & 37.2 & $< $2.3 & $<$3.3 & 2 \\
\pga & 1998 Nov 12 & 70.1 & 64.8 & 1.3$\pm$0.5 & $<$1.0 & 3 \\
\pgb & 1999 Nov 12 & 38.0 & 40.1 & $<$1.4 & $<$1.3 &  1 \\
\mrk  &  1994 Dec 05 & 19.4 & 18.2 & 9.9$\pm$1.2 & $7.4\pm$1.2  & 2 \\
\mrk &  1999 Nov 10 & 86.4 & 90.0 & 11.2$\pm$0.6 &  9.1$\pm$0.5 &  4 \\
\pge & 1998 Mar 24 & 18.3 & 14.7 & $<$2.1 & $<$2.1 & 2\\

\pgf &  1999 Oct 30 &  31.1 &  28.4 & 20.2$\pm$1.2 & 13.3$\pm$1.0  & 5 \\
\enddata

\tablecomments{The SIS count rates are for the 0.6$-$9.5 keV band extracted from a circular region of radius 3.2\arcmin.  
Count rate upper limits are at the 3$\sigma$ significance level. The count rates and exposure times
for all the observations with exception of the one of  \irasa\ have been obtained selecting source and background regions from the cleaned event files of the NASA Tartarus database  (version 3.1; http://astro.ic.ac.uk/Research/Tartarus). The upper limits and exposure times for the observation of \irasa\ have been obtained from~\cite{1999ApJ...519..549G}.}

\tablenotetext{a}{The exposure times and photon counts are obtained after screening the data.}

\tablenotetext{b}{{\sc references:} (1)~this work; (2)~\cite{1999ApJ...519..549G}; (3)~\cite{2000ApJ...533L..79M}; (4)~\cite{2000ApJ...545L..23M} (5)~\cite{2001ApJ...546..795G}.}


\end{deluxetable*}
 
\subsection{\asca\ observations}

For the \asca\ observations of our sample we used {\sc xselect} to extract the spectra from the  previously processed clean events files publicly available in the Tartarus (Version 3.1) database.\footnote{http://tartarus.gsfc.nasa.gov} We have extracted spectra from only the SIS cameras. This was based on the fact that  the SIS cameras have a wider range of energy sensitivity and better angular resolution than the GIS proportional counters. For the SIS, the counts from each quasar were extracted from circular regions with aperture radii of 3.2\arcmin.   The background spectra for the SIS observations were extracted from  source-free polygons covering most of the nominal chip at distance $>4.2$\arcmin\ from the source center. The background subtracted SIS count rates in  the 0.6$-$9.5~keV band are presented in Table~\ref{tab:asca}.

\subsection{\sax\ and \einstein\ observations}
For the \sax\ and \einstein\ observations of our sample we collected  the count rates from  already published works. 
The count rates for the \sax\ MECS observations were obtained from \cite{2004A&A...420...79B} in the case of \mrk\ and from  \cite{1999ApJ...525L..69B} in the case of \cso. The published count rates were extracted in the 1.8--10~keV band using circular regions with radii 2\arcmin\ and 3\arcmin\ for \mrk\ and \cso, respectively.
The count rates for the \einstein\ IPC observations of our sources were obtained from \cite{1994ApJS...92...53W}. These count rates were extracted in the 0.16--3.5~keV band using circular regions with 3\arcmin\ radii. 
We used the Portable, Interactive Multi-Mission Simulator software
 \citep[PIMMS;][]{1993Legac...3...21M} to derive X-ray fluxes from the published
\sax\ and \einstein\ count rates.

\subsection{New \XR\ detection of \irasb}
In this work we report for the first time an \XR\ detection of \irasb,  a highly polarized BAL quasar that has been studied in detail at other wavelengths \citep[e.g.,][]{1992ApJ...397..442B, 2001ApJ...563..512H}. We detect this source with a statistical significance of $\approx 4 \sigma$ in the MOS2 observation performed on 2005 Nov 26, $\approx 3 \sigma$ in the MOS1 observation performed on 2006 Jan 16, and $\approx  5 \sigma$ in our new ACIS-S observation performed on 2010 Jul 28. The significances of the detections of \irasb\ were calculated by estimating the Poisson probability of obtaining at least the observed number of counts based on the expected background. This probability provides a statistical test of the null hypothesis that the observed photon counts solely originate from the background.
For the MOS2 observation performed on 2005 Nov 26 there are 21 counts detected with a predicted background number of 6.63, and therefore the null probability is $6.6 \times 10^{-6}$; i.e., we reject the null hypothesis with a significance of $\approx 4.4\sigma$. For the MOS1 observation performed on 2006 Jan 16 there are 19 counts detected with a predicted number of 7.45, and therefore the null probability is $2.8 \times 10^{-4}$  (significance of $\approx 3.4\sigma$).  For the \chandra\ observation there are 6 counts detected with a predicted background number of 0.22, and therefore the null probability is $1.3 \times 10^{-7}$  (significance of $\approx 5.1\sigma$).
The detections of \irasb\  in the MOS cameras are close to the detection
threshold, and therefore different choices of energy range and/or
source and background extraction regions could result in a
non-detection of the source as in the case of  \cite{2007A&A...471..775R}. For the observations of \irasb\  we have used our default annular background  region. As an additional test of the detections of \irasb\ in the MOS cameras, we select  two  80\arcsec\ radius \hbox{source-free} circular regions in the vicinity of the source  as the background. Using these background regions for the MOS2 observation performed on 2005 Nov 26, the predicted numbers of background counts are 7.98 and 9.66. Thus the detection significances are $\approx 3.7\sigma$ and $\approx 3.1\sigma$. For the MOS1 observation performed on 2006 Jan 16  the  predicted numbers of background counts are 6.41 and 7.76; therefore the detection significances are $\approx 3.9\sigma$ and $\approx 3.3\sigma$. The detection of \irasb\ in its \chandra\ observation has $\approx 5\sigma$ significance independent of the selection of the background region. For example,  using as background regions two 20\arcsec\ radius \hbox{source-free} circular regions in the vicinity of the source, the expected numbers of background counts are 0.24 and 0.32. Thus, the detection significances are $\approx 5.1\sigma$ and $\approx 4.7\sigma$. 

\begin{deluxetable*}{cccccccc}
\tablecolumns{8}
\tabletypesize{\scriptsize}
 \tablewidth{0pt}
\tablecaption{Best-Fit Power-Law \hbox{X-ray} Spectral Parameters \label{tab:xfit}}
\tablehead
{
\colhead{} & \colhead{} & \colhead{\nh  \tablenotemark{a}} & \colhead{} & \colhead{}  \\
\colhead{\sc object name} & \colhead{\sc obs. date} & \colhead{(10$^{22}$~cm$^{-2}$)} & \colhead{$\Gamma$} & \colhead{$C$-stat(DOF)} & \colhead{Instrument}
}
\startdata

\irasa & 1991 Mar 17 & $<$0.15 & 2.71 &  22.0(22) & PSPC  \\
& 2000 Mar 21 & $<$0.52 & 2.71 &  7.0(8) & ACIS-S  \\ 
&  2001 Oct 24 & $<$0.06 & 2.71$\pm$0.19 &  301.3(345) & EPIC  \\
 & 2010 Jun 18 & $<$0.12 & 2.71 &  36.4(27) & ACIS-S  \\ 

 \vspace{5pt} \\

\pga   &  1998 Nov 12 & $12^{+10}_{-6}$ & 1.48 &  303.7(321) & SIS   \\
& 2010 Jan 11 & $7.3_{-5.3}^{+7.1}$ & 1.48$\pm$0.86 &  58.2(52) & ACIS-S  \\

\vspace{5pt}  \\

\pgb  & 1992 May 17 & $<$0.19 & 2.04 &  11.5(9) & PSPC  \\
&  2003 May 04 & $<$0.01& 2.04$\pm$0.15 &  557.0(325) & EPIC  \\ 
& 2010 Jan 11 & $<$0.27 & 2.04 &  4.3(11) & ACIS-S  \\ 
 
 \vspace{5pt}  \\

\pgc &  2003 May 04 & $<$0.01 & 1.39$\pm$0.04 &  1204.4(1339) & EPIC  \\
&  2005 Jan 05 & $<$0.15 & 1.03$\pm$0.06 &  521.8(545) & ACIS-S  \\ 
  
\vspace{5pt}  \\

\Qa 
& 1991 Dec 21 & $<$8.1 & 1.58 &  22.3(28) & PSPC  \\
& 1993 Jan 05 & $<$3.6 & 1.58 &  15.5(12) & PSPC  \\
& 1993 Jun 22 & $<$2.6 & 1.58 &  39.0(66) & PSPC  \\
&  2000 Feb 08 & $3.8_{-2.8}^{+3.5}$ & 1.58 &  26.4(36) & ACIS-S  \\ 
&   2001 Jul 11 & $0.67_{-0.55}^{+0.69}$ & 1.83$\pm$0.20 &  441.8(584) & EPIC  \\ 
&  2004 Jul 15 & $<$1.2 & 1.40$\pm$0.17 &  542.9(606) & EPIC  \\

\vspace{5pt}  \\

\mrk & 1991 Jun 7 & $<$0.01 & 1.03 &  121.4(103) & PSPC  \\ 
& 1994 Dec 05 & $<$0.10 & 1.00$\pm$0.22 &  251.4(304) & SIS  \\
& 1999 Nov 10 &  $<$0.02 & 0.94$\pm$0.10 & 814.8(882)  & SIS   \\
& 2000 Oct 19 & $<$0.00 & 1.03$\pm$0.06 &  731.8(629) & ACIS-S  \\ 
&  2001 Jun 07 & $<$0.00 & 1.06$\pm$0.04 &  1919.7(1663) & EPIC  \\ 
&  2003 Feb 03 & $<$0.00 & 1.07$\pm$0.06 &  719.9(603) & ACIS-S  \\
& 2003 Feb 11 & $<$0.00 & 1.05$\pm$0.06 &  847.9(582) & ACIS-S  \\
& 2003 Feb 20 & $<$0.00 & 0.95$\pm$0.07 &  694.0(605) & ACIS-S  \\
& 2010 Jul 11 & $<$0.03  & 0.89$\pm$0.20 &  189.2(211) & ACIS-S  \\ 

 \vspace{5pt}  \\

\irasb  &  2005 Nov 26 & $<$4.5 & 1.80 &  24.2(18) & EPIC  \\ 
&  2006 Jan 16 & $<$1.5 & 1.80 &  30.3(18) & EPIC  \\
& 2010 Jul 28 & $<$113 & 1.80 &  8.6(5) & ACIS-S  \\ 
  
\vspace{5pt}  \\

\cso &  2001 Jul 30 & $<$3.4 & $1.50_{-0.32}^{+0.42}$ &  47.8(63) & EPIC  \\ 
&  2001 Dec 08 & 1.20$\pm$0.30 & 1.83$\pm$0.07 &  1036.9(1161) & EPIC  \\
&  2001 Dec 13 & $2.8_{-1.5}^{+1.9}$ & 2.02$\pm$0.29 &  213.1(238) & EPIC  \\ 
& 2010 Mar 21 & $<$5.3 & 1.97$\pm$0.46 &  79.0(82) & ACIS-S  \\
 
 \vspace{5pt}  \\

\SBS & 1993 Ago 14 & $<$4.5 & 1.74 &  13.5(24) & PSPC  \\
& 2000 Mar 22 & $<$4.0 & 1.55$\pm$0.49 &  56.6(65) & ACIS-S  \\ 
&  2002 Feb 03 & $<$1.6 & 1.73$\pm$0.26 &  214.7(254) & EPIC  \\
&  2002 Feb 06 & $0.52_{-0.37}^{+0.41}$ & 1.76$\pm$0.13 &  576.9(647) & EPIC  \\ 

\vspace{5pt}  \\

\pge  &  2009 Dec 19 & $<$0.04 & 1.77 &  49.2(44) & EPIC  \\
&  2009 Dec 31 & $<$0.08 & 1.77 &  141.2(124) & EPIC  \\
&  2010 Jan 02 & $<$0.12  & $1.77_{-0.30}^{+0.33}$ &  191.0(195) & EPIC  \\ 
& 2010 Jun 26 & $<$0.11 & 1.77 &  19.0(13) & ACIS-S  \\
 
 \vspace{5pt}  \\

\pgf  & 1991 Nov 30 & $<$0.20 & 0.91 &  39.9(46) & PSPC  \\ 
 & 1999 Oct 30  & 1.14$\pm$0.50 & 1.91$\pm$0.26 & 506.1(599) & SIS  \\
&  2002 Sep 01 & $<$0.03 & 0.84$\pm$0.09 &  390.8(399) & ACIS-S  \\ 
&  2003 May 14 & $<$0.02 & 1.59$\pm$0.06 &  896.1(958) & EPIC  \\ 
&  2005 Nov 20 & $<$0.00 & 0.84$\pm$0.08 &  1095.1(1023) & EPIC  \\
&  2007 May 03 & $<$0.02 & 0.66$\pm$0.11 &  583.6(585) & EPIC  \\ 
&  2007 May 19 & $<$0.01 & 0.69$\pm$0.07 &  1053.0(1164) & EPIC  \\
&  2007 May 21 & $<$0.01  & 0.62$\pm$0.06 &  1495.4(1509) & EPIC  \\
&  2007 Nov 05 & $<$0.02 & 0.58$\pm$0.09 &  779.6(891) & EPIC  \\ 

\enddata
\tablecomments{
The best-fit  column densities  were obtained from a \PL\ model  with intrinsic absorption at the redshift of the source. The same model was used to estimate the best-fit  photon index and $C$-statistic for  the  observations of \pga, \Qa, \SBS, \cso, and \pgf. The best-fit  photon index and $C$-statistic for all the other observations were obtained from an unabsorbed  \PL\ model. In all the fits we have assumed Galactic absorption with column densities obtained from \cite{1990ARA&A..28..215D}.
Errors and upper limits represent 90\% confidence intervals for each value,
taking one parameter to be of interest \citep[$\Delta C$=2.71; e.g.,][]{1976ApJ...210..642A}. }

\tablenotetext{a}{Intrinsic neutral hydrogen column density in units of 10$^{22}$~cm$^{-2}$.
}
\end{deluxetable*}

\begin{figure*}
   \includegraphics[width=16cm]{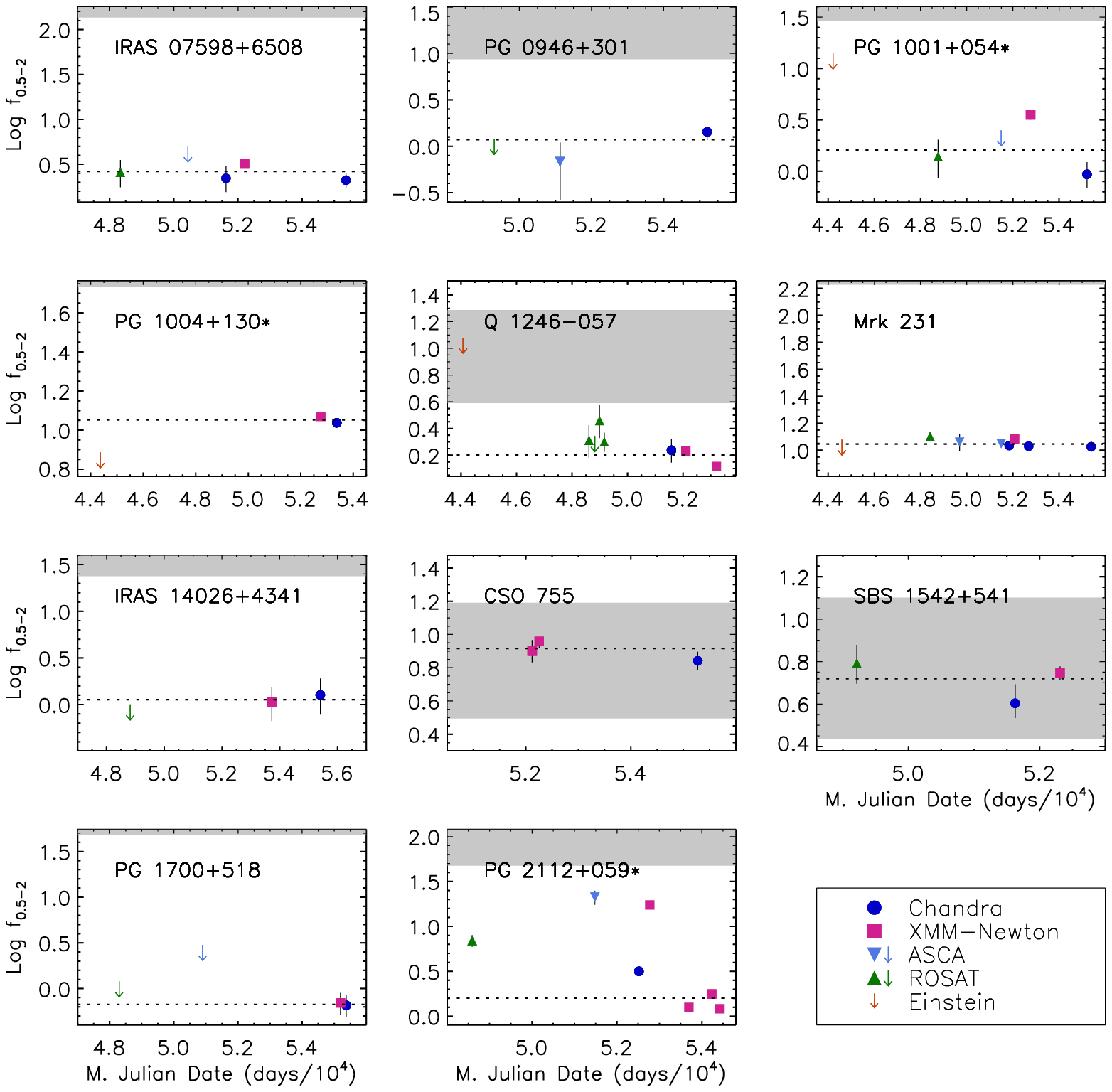}
        \centering
       \caption{Logarithm of the observed flux in the 0.5$-$2~keV band (in units of $10^{-14}$~\flux) versus modified Julian date for the BAL quasars in our sample. The circles, squares, inverted triangles, and triangles indicate observations performed with \chandra, \xmm, \asca, and \rosat, respectively.  Given that \mrk\ did not present variability in the set of \chandra\ observations performed in 2003, for visualization purposes we have plotted these observations as a single point with their mean flux  and a typical error bar.  The same procedure has been applied to the \xmm\ observations of \irasb,  the \xmm\ observations of \cso\ performed in 2001 Dec, the \xmm\ observations of \SBS, the \xmm\ observations of \pge, and the \xmm\ observations of \pgf\ performed in 2007 May.  The upper limits at early times found in the panels of \pgb, \pgc, \Qa, and \mrk\ correspond to the \einstein\ observations. The shaded areas in each panel correspond to the expected range of X-ray fluxes obtained assuming a typical \aox\ \citep[see Table 5 of][]{2006AJ....131.2826S} and a \PL\ spectrum with $\Gamma=1.9$. The sources marked with * present evidence of variability  in at least one of the energy bands considered. The horizontal dotted lines correspond to the best-fit constant flux.}
     \label{fig:var1}
     \end{figure*}

\begin{figure*}
   \includegraphics[width=16cm]{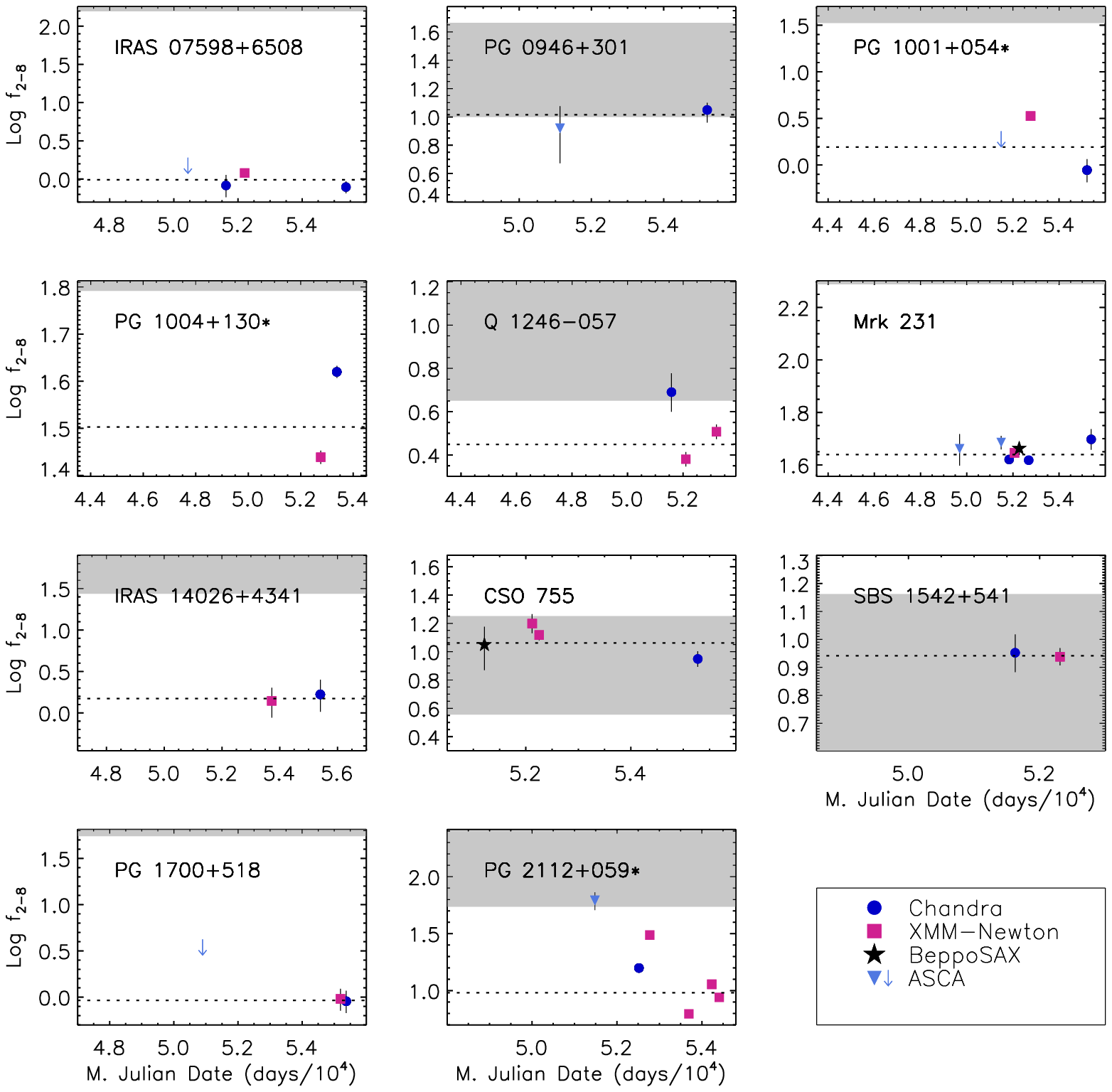}
        \centering
    \caption{Logarithm of the observed flux in the 2$-$8~keV band (in units of $10^{-14}$~\flux) versus modified Julian date for the BAL quasars in our sample. The circles, squares, stars and inverted triangles indicate observations performed with \chandra, \xmm, \sax, and \asca, respectively.   See the Figure~\ref{fig:var1} caption for additional details about the presentation.}
     \label{fig:var2}
     \end{figure*}
     
     \begin{figure*}
   \includegraphics[width=14cm]{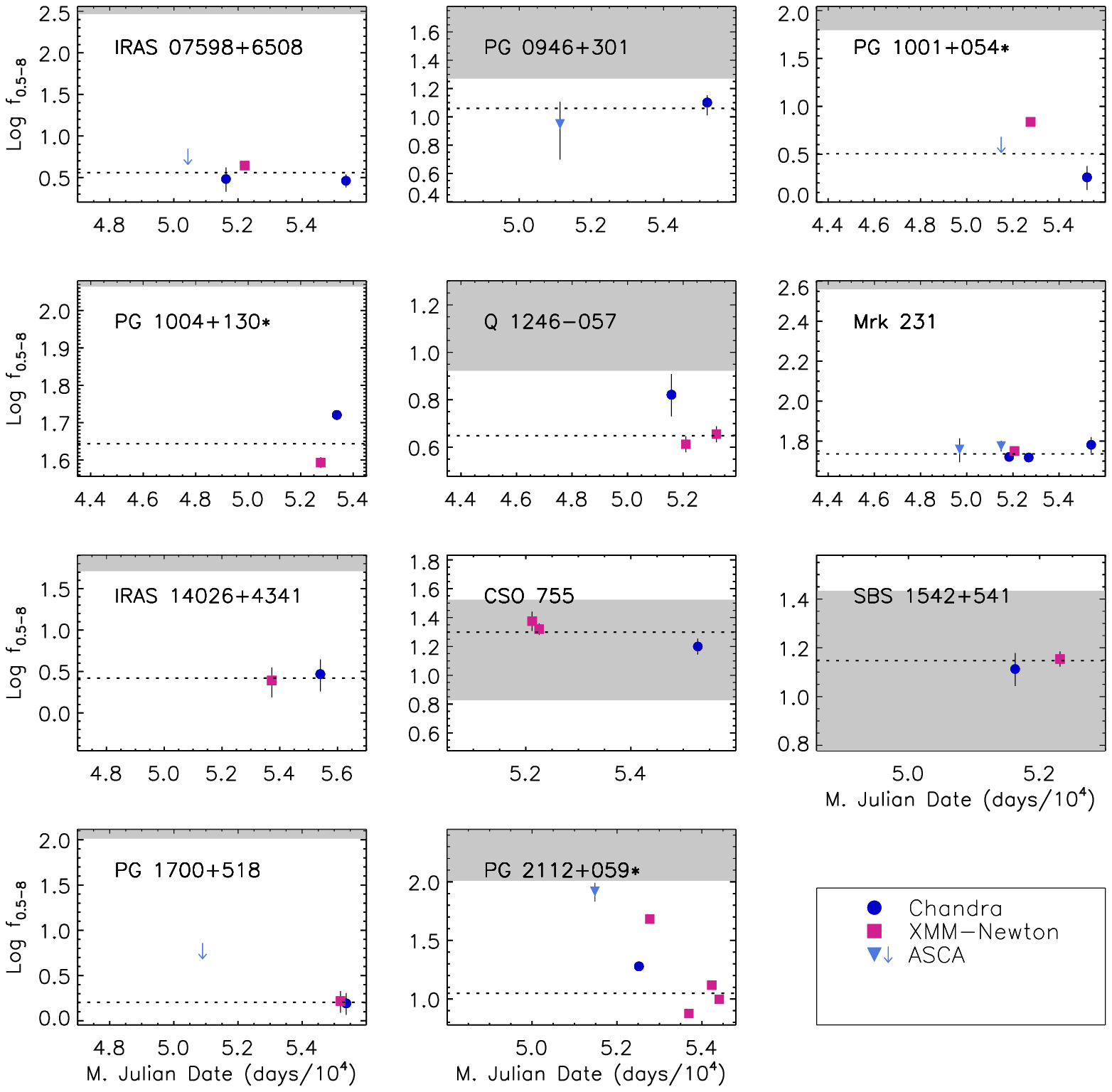}
        \centering
    \caption{Logarithm of the observed flux in the 0.5$-$8~keV band (in units of $10^{-14}$~\flux) versus modified Julian date for the BAL quasars in our sample. The circles, squares, and inverted triangles indicate observations performed with \chandra, \xmm, and \asca, respectively.   See the Figure~\ref{fig:var1} caption for additional details about the presentation.}
     \label{fig:var3}
     \end{figure*}

\section{SPECTRAL ANALYSIS} \label{S:span}

The spectral analysis of our sample was undertaken with the main goal  of searching for long-term flux variability within the set of \XR\ observations  of each source. Our objective is not to perform a detailed spectral modeling of each source in our sample but to obtain robust basic flux measurements.
Spectral analysis was performed using \xspec\ v.12.0; the fitting bands used were \FB\ for the \chandra\ observations, \hbox{0.5--10~keV} for the \xmm\ MOS observations, \hbox{0.3--10~keV} for the \xmm\ pn observations, \hbox{0.6--9.5~keV} for the \asca\ SIS observations, and \SB\ for the \rosat\ PSPC observations. 
The spectral fits were performed in the full fitting band of each
instrument using the $C$-statistic \citep{1979ApJ...228..939C}. 
  The $C$-statistic may not be appropriate to utilize when performing background subtraction. To use the $C$-statistic properly, we fit the background spectra with a flat response using the {\sc cplinear} model.\footnote{This model is included in the {\sc acis-extract} package \citep{2010ApJ...714.1582B}. } The background model is scaled and subtracted when we fit the source spectra.

For  \chandra\  we performed fits using each of the new and archival observations. For  \xmm\  we  performed fits using each of the EPIC cameras with detected spectra separately; we also performed joint fits using the three EPIC cameras together. The fits from each separate EPIC camera are consistent with the joint fits, and therefore we present only the joint fits in this work.  The same conclusion was obtained for the ASCA SIS cameras, and therefore we present only the joint fits of the SIS0 and SIS1 cameras in this work.

For deriving fluxes and luminosities of each source we used a simple \PL\ model including intrinsic absorption in the cases where significant absorption was found ($>99$\% level of significance); in all cases we included Galactic absorption from \cite{1990ARA&A..28..215D}. Galactic absorption column densities  \cite[from][]{1990ARA&A..28..215D} were obtained at the optical coordinates of the sources (see Tables 1 and 2) through the use of the HEASARC \nh\ tool.\footnote{http://heasarc.gsfc.nasa.gov/cgi-bin/Tools/w3nh/w3nh.pl} 
The \PL\ photon index ($\Gamma$) was free to vary for those \chandra, \xmm, and \asca\ observations where the \hbox{signal-to-noise} ratio (S/N) of the observation was greater than seven. The value of S/N is estimated from the total number of counts divided by the average of the upper and lower error bars; the total photon counts with their error bars are found in Tables~\ref{tab:chao}, \ref{tab:xmmo} and \ref{tab:asca}.   In the cases where ${\rm S/N} \leq 7$ and for the \rosat\ observations,\footnote{The \SB\ band observations with \rosat\  cannot be used to give a reliable estimation of the full-band (\FB) photon index.}  we fixed the  photon index to the weighted average if there are other observations of the source with ${\rm S/N} >7$ or otherwise to 1.8.   It is important to emphasize that we are not correcting for intrinsic absorption when we derive fluxes.  This allows us to make reliable estimations of the absorbed fluxes and luminosities regardless of the model that we are using since these quantities are rather insensitive to spectral model choice. Given the low S/N of most of the spectra analyzed in this work,  a simple or absorbed \PL\ gives an acceptable fit for almost every  observation;  however, in a few cases of high S/N ($\rm S/N>25$)  where we find evidence of complexity in the spectra, we also estimate fluxes and luminosities using more complex models (see \S\ref{S:vaCM}).

The best-fitted parameters using a \PL\ model are presented in Table~\ref{tab:xfit}. 
There are five sources that presented evidence of absorption in their spectra; these sources are \pga, \Qa, \cso, \SBS, and \pgf\ (see  Table~\ref{tab:xfit}).  \pga\ shows in both \chandra\ and \asca\ observations \hbox{moderate-to-strong} absorption with column densities $\sim 10^{23}~\cmsq$. 
This source presents moderate and low S/N in its \chandra\ and \asca\ observations, respectively. Therefore, new \chandra\ and/or \xmm\ observations with exposures $\gtrsim 20$~ks   are needed  in order to better constrain the absorption found in \pga.
\Qa\ shows moderate intrinsic absorption ($\nh\sim 10^{22}~\cmsq$) as we confirm in the \chandra\ observation performed in 2000 and the \xmm\ observation performed in 2001 \citep[first studied by][]{2003AJ....126.1159G}.
\cso\ presents moderate intrinsic absorption ($\nh\sim 10^{22}~\cmsq$) as we confirm in the \xmm\ observations \citep[first studied by][]{2005AJ....130.2522S}. 
\SBS\ shows low intrinsic absorption ($\nh \lesssim 10^{22}~\cmsq$) as we confirm in the last \xmm\ observation performed in 2002 \citep[first studied by][]{2003AJ....126.1159G}. 
\pgf\  presents evidence of moderate absorption only in its \asca\ observation performed in October 1999 \citep[first studied by][]{2001ApJ...546..795G}; however, the flat measured power laws in other observations suggest absorption is likely present.

The derived \XR\ fluxes in the \SB, \HB, and \FB\ bands from simple or absorbed \PL\ fits are presented in Table~\ref{tab:flu} and Figures~\ref{fig:var1}, \ref{fig:var2}, and \ref{fig:var3}. There we have also presented fluxes from observations with \sax\ and \einstein; these fluxes were obtained from published count rates using the PIMMS tool (see \S\ref{S:data}). We assumed \PL\ spectra with the average parameters ($\Gamma$ and \nh) taken from spectral fits of the high S/N observations of each source. The error bars for the \sax\ MECS observations were obtained by propagating the uncertainties on the instrumental background\footnote{To estimate the background count rates we used the background files found at the NASA HEASARC site (http://heasarc.nasa.gov/docs/sax/archive/calibration\_files.html).}  and source count rates presented in \cite{1999ApJ...525L..69B} and \cite{2004A&A...420...79B}. The \einstein\ observations of our sources provided only upper limits. We confirmed that the estimates of \einstein\ fluxes with PIMMS give a reliable result by comparing our values with those of \cite{1994ApJS...92...53W}; in general the differences are less than 10\%. Similarly, our \sax\ fluxes are consistent within 10\% with those of \cite{1999ApJ...525L..69B} and \cite{2004A&A...420...79B}.

Note that we have only estimated fluxes and luminosities in the bands where  the instruments are sensitive. Therefore, in the case of the \rosat\ PSPC and \einstein\ IPC observations we only present fluxes in the \SB\ band, and in the case of the \sax\ MECS we only present fluxes in the \HB\ band. 

 \begin{deluxetable*}{ccccccc}
\tablecolumns{8}
\tabletypesize{\scriptsize}
 \tablewidth{0pt}
\tablecaption{Derived X-ray Fluxes \label{tab:flu} from Power-Law Fits}
\tablehead
{
\colhead{\sc object name\tablenotemark{a}} & \colhead{\sc obs. date} & \colhead{$f_{0.5-2}$\tablenotemark{b}} & \colhead{$f_{2-8}$\tablenotemark{b}} & \colhead{$f_{0.5-8}$\tablenotemark{b}} & \colhead{$f_{\rm 2keV}$\tablenotemark{c}} & \colhead{Instrument}
}
\startdata
{ \irasa} 
& 1991 Mar 17 & 2.60$\pm$0.89  & ... & ...& 2.88$\pm$0.99 & PSPC \\ 
& 1996 Dec 21 & $<$5.0 & $<$1.9 & $<$7.0 & $<$5.6   &     SIS \\
& 2000 Mar 21 & $2.21_{-0.66}^{+0.82}$ & $0.83_{-0.25}^{+0.31}$ & $3.03_{-0.90}^{+1.13}$ & $2.45_{-0.73}^{+0.91}$ & ACIS-S  \\
& 2001 Oct 24 & 3.20$\pm$0.29 & 1.20$\pm$0.11 & 4.40$\pm$0.39 & 3.55$\pm$0.32  & EPIC  \\
& 2010 Jun 18 & 2.10$\pm$0.38 & 0.79$\pm$0.14 & 2.89$\pm$0.52 & 2.33$\pm$0.42 & ACIS-S 
   \vspace{0.0pt}  \\

{\pga}  
  & 1993 Nov 14 &  $<$1.2 & ... & ... &  $<$0.8 & PSPC  \\
 
 & 1998 Nov 12 &  0.68$\pm$0.42 & 8.3$\pm$3.6 & 8.9$\pm$3.9  & $<$0.6  & SIS \\
 & 2010 Jan 11 & $1.43_{-0.27}^{+0.18}$ & $11.2_{-2.1}^{+1.4}$ & $12.6_{-2.3}^{+1.6}$ & 0.69$\pm$0.11 & ACIS-S  \\
  
  \vspace{0.0pt}  \\

{\pgb*}
  & 1979 Dec 02  & $<$14 & ... & ... & $<$24 &  IPC\\
 & 1992 May 17 & $1.40_{-0.52}^{+0.64}$  & ... & ...& $2.4_{-0.9}^{+1.1}$ & PSPC \\
 & 1999 Nov 12 & $<$2.5 & $<$2.3 & $<$4.8 & $<$ 4.2 & SIS \\
 & 2003 May 04 & 3.53$\pm$0.36 & 3.35$\pm$0.34 & 6.88$\pm$0.70 & 5.99$\pm$0.61  & EPIC  \\
 & 2010 Jan 11 & 0.93$\pm$0.27 & 0.88$\pm$0.25 & 1.81$\pm$0.52 & 1.58$\pm$0.45 & ACIS-S  \\
 
 \vspace{5pt}  \\

{\pgc*} 
& 1980 May 09 &    $<$7.7       & ... & ... & $<$20 & IPC \\
 & 2003 May 04 & 11.76$\pm$0.38 & 27.45$\pm$0.89 & 39.2$\pm$1.3 & 28.27$\pm$0.92  & EPIC  \\ 
 & 2005 Jan 05 & 10.90$\pm$0.32 & 41.7$\pm$1.2 & 52.6$\pm$1.6 & 29.72$\pm$0.88 & ACIS-S  \\

 \vspace{0.0pt}  \\ 
 
 {\Qa}
& 1979 Jul 20 & $<$12 & ... & ... & $<$38 & IPC \\
 
 & 1991 Dec 21 & 2.07$\pm$0.57  & ... & ...& 4.4$\pm$1.2 & PSPC \\
& 1992 Jul 17  &  $<$2.2 & ... & ...&   $<$6.2&  PSPC \\ 
& 1993 Jan 05 &  3.18$\pm$0.91  & ... & ...& 12.5$\pm$3.6 & PSPC  \\ 
& 1993 Jun 22 & 2.21$\pm$0.36  & ... & ...& 7.5$\pm$1.2 & PSPC \\
& 2000 Feb 08 & 1.73$\pm$0.35 & 4.9$\pm$1.0 & 6.6$\pm$1.4 & 1.97$\pm$0.41 & ACIS-S  \\
& 2001 Jul 11 & 1.70$\pm$0.13 & 2.40$\pm$0.18 & 4.10$\pm$0.31 & 6.28$\pm$0.48  & EPIC  \\ 
& 2004 Jul 15 & 1.30$\pm$0.10 & 3.22$\pm$0.25 & 4.52$\pm$0.35 & 3.94$\pm$0.30  & EPIC  \\

 \vspace{0.0pt}  \\

{\mrk} 
 & 1980 Dec 23  & $<$12 &  ... & ... & $<$33 & IPC \\
& 1991 Jun 07 & 12.73$\pm$0.93  & ... & ...& 34.6$\pm$2.5 & PSPC \\

& 1994 Dec 05 & 11.5$\pm$1.6 &  45.9$\pm$6.3 &  57.3$\pm$7.8 & 31.6$\pm$4.3 & SIS \\
& 1999 Nov 10 & 11.2$\pm$0.67  &  48.5$\pm$2.9 & 59.6$\pm$3.6 & 31.7$\pm$1.9  & SIS   \\
& 2000 Oct 19 & 10.83$\pm$0.34 & 41.7$\pm$1.3 & 52.6$\pm$1.7 & 29.48$\pm$0.93 & ACIS-S  \\
& 2001 Jun 07 & 12.07$\pm$0.35 & 44.3$\pm$1.3 & 56.3$\pm$1.6 & 32.26$\pm$0.94  & EPIC  \\ 
 & 2001 Dec 29 & ... &  46.0$\pm$2.5 & ... & ... & MECS \\
 & 2003 Feb 03 & 10.98$\pm$0.33 & 39.7$\pm$1.2 & 50.7$\pm$1.6 & 29.21$\pm$0.89 & ACIS-S  \\
 & 2003 Feb 11 & 10.88$\pm$0.34 & 40.7$\pm$1.3 & 51.6$\pm$1.6 & 29.29$\pm$0.91 & ACIS-S  \\
 & 2003 Feb 20 & 10.29$\pm$0.33 & 44.1$\pm$1.4 & 54.4$\pm$1.8 & 29.09$\pm$0.95 & ACIS-S  \\
 & 2010 Jul 11 & 10.62$\pm$0.96 & 49.8$\pm$4.5 & 60.5$\pm$5.5 & 40.0$\pm$2.8 & ACIS-S  \\
  \vspace{0.0pt}   \\
 
{\irasb} 
  & 1992 Jul 18 &  $<$1.0 & ... & ... &  $<$2.1 & PSPC  \\
& 2005 Nov 26 & 1.28$\pm$0.44 & 1.69$\pm$0.58 & 3.0$\pm$1.0 & 2.73$\pm$0.94  & EPIC  \\
& 2006 Jan 16 & 0.84$\pm$0.42 & 1.11$\pm$0.55 & 1.95$\pm$0.97 & 1.79$\pm$0.89  & EPIC  \\
 & 2010 Jul 28 & $1.27_{-0.48}^{+0.64}$ & $1.67_{-0.64}^{+0.84}$ & $2.9_{-1.1}^{+1.5}$ & $2.7_{-1.0}^{+1.4}$ & ACIS-S 
  \vspace{0.0pt}   \\

{\cso} 
 & 1999 Feb 02 & ... &  11.2$\pm3.8$ & ... & ... &  MECS \\
 & 2001 Jul 30 & 8.0$\pm$1.2 & 15.8$\pm$2.4 & 23.7$\pm$3.7 & 32.5$\pm$5.0  & EPIC  \\
 & 2001 Dec 08 & 9.09$\pm$0.29 & 13.10$\pm$0.42 & 22.18$\pm$0.71 & 31.1$\pm$1.0  & EPIC  \\ 
 & 2001 Dec 13 & 8.48$\pm$0.84 & 11.1$\pm$1.1 & 19.6$\pm$1.9 & 18.8$\pm$1.9  & EPIC  \\
 & 2010 Mar 21 & 6.94$\pm$0.87 & 8.9$\pm$1.1 & 15.8$\pm$2.0 & 20.5$\pm$2.6 & ACIS-S  \\
  \vspace{0.0pt}   \\
 
{\SBS}
& 1993 Ago 14 & 6.2$\pm$1.3  & ... & ...& 15.4$\pm$3.2 & PSPC \\ 
& 2000 Mar 22 & 4.01$\pm$0.62 & 9.0$\pm$1.4 & 13.0$\pm$2.0 & 10.3$\pm$1.6 & ACIS-S  \\
& 2002 Feb 03 & 5.34$\pm$0.50 & 8.55$\pm$0.80 & 13.9$\pm$1.3 & 19.6$\pm$1.8  & EPIC  \\
& 2002 Feb 06 & 5.81$\pm$0.29 & 8.79$\pm$0.44 & 14.60$\pm$0.73 & 22.1$\pm$1.1  & EPIC  \\

\vspace{0.0pt}   \\

{\pge} 
   & 1991 Feb 09 & $<$1.2 & ... & ... &  $<$2.5 & PSPC  \\
& 1998 Mar 24 & $<$3.0 & $<$4.2 & $<$7.2 & $<$6.4 & SIS \\
& 2009 Dec 19 & 0.83$\pm$0.19 & 1.15$\pm$0.26 & 1.98$\pm$0.45 & 1.77$\pm$0.40  & EPIC  \\
 & 2009 Dec 31 & 0.50$\pm$0.14 & 0.68$\pm$0.20 & 1.18$\pm$0.34 & 1.06$\pm$0.30  & EPIC  \\
 & 2010 Jan 02 & 0.76$\pm$0.16 & 1.04$\pm$0.21 & 1.80$\pm$0.37 & 1.61$\pm$0.33  & EPIC  \\ 
& 2010 Jun 26 & 0.66$\pm$0.18 & 0.90$\pm$0.25 & 1.56$\pm$0.43 & 1.39$\pm$0.38 & ACIS-S  \\
\vspace{0.0pt}   \\

{\pgf*} 
& 1991 Nov 30 & 7.0$\pm$1.0  & ... & ...& 19.6$\pm$2.9 & PSPC  \\ 

& 1999 Oct 30  &   21.2$\pm$3.8 & 62$\pm$11 &  83$\pm$15 & 77$\pm$14 & SIS \\

& 2002 Sep 01 & 3.17$\pm$0.14 & 15.81$\pm$0.69 & 18.98$\pm$0.83 & 8.94$\pm$0.39 & ACIS-S  \\ 
& 2003 May 14 & 17.35$\pm$0.74 & 30.8$\pm$1.3 & 48.1$\pm$2.1 & 42.6$\pm$1.8  & EPIC  \\
& 2005 Nov 20 & 1.25$\pm$0.07 & 6.26$\pm$0.34 & 7.51$\pm$0.40 & 3.54$\pm$0.19  & EPIC  \\ 
& 2007 May 03 & 1.88$\pm$0.13 & 11.98$\pm$0.86 & 13.86$\pm$0.99 & 5.42$\pm$0.39  & EPIC  \\
& 2007 May 19 & 1.67$\pm$0.08 & 10.26$\pm$0.47 & 11.93$\pm$0.55 & 4.81$\pm$0.22  & EPIC  \\
& 2007 May 21 & 1.74$\pm$0.06 & 11.86$\pm$0.44 & 13.60$\pm$0.51 & 5.04$\pm$0.19  & EPIC  \\
& 2007 Nov 05 & 1.21$\pm$0.07 & 8.73$\pm$0.52 & 9.94$\pm$0.59 & 3.53$\pm$0.21  & EPIC 

\enddata


\tablecomments{Fluxes are obtained from the best-fitting parameters for a simple
power-law model (most cases) or an intrinsic absorbed power-law model
(for observations as indicated in the previous table).  In both cases,
fixed Galactic absorption was also included. The fluxes are not corrected for intrinsic absorption.}


\tablenotetext{a}{The sources marked with * present evidence 
of variability in at least one of the energy bands considered.}

\tablenotetext{b}{Observed fluxes in the  0.5$-$2, 2$-$8, and 0.5$-$8~keV  bands in units of $10^{-14}$~\flux. }

\tablenotetext{c}{Observed flux densities at \RF\  2~keV in units of $10^{-32}$~\flux Hz$^{-1}$. }

\end{deluxetable*}

\section{RESULTS AND DISCUSSION}  \label{S:resu}
In this section we describe the main results found in the \XR\ variability study of our BAL quasar sample. In \S\ref{S:vaPL} we constrain variability using  simple \PL\ models (as described in \S\ref{S:span}) to estimate fluxes; in \S\ref{S:vaCM} we constrain variability  using more complex models to estimate fluxes; finally, in \S\ref{S:vaan} we examine possible physical scenarios to explain the results found.

\begin{deluxetable*}{cccccccccc}
\setlength{\tabcolsep}{0.01in}
\tablecolumns{10}
\centering
\tabletypesize{\scriptsize}
 \tablewidth{0pt}
\tablecaption{Tests for Variability: Constant Flux Fits \label{tab:fave}}
\tablehead
{
\colhead{} &
\multicolumn{3}{c}{\underline{Fluxes~$[10^{-14}~\flux]$}\tablenotemark{a}} & \multicolumn{3}{c}{\underline{\hspace{40pt}$\chi^2$/{\sc dof}\hspace{40pt}}} & \multicolumn{3}{c}{\underline{\hspace{20pt} \sc \% significance \hspace{20pt}}}  \\
 \colhead{\sc object name} & \colhead{0.5$-$2~keV} & \colhead{2$-$8~keV} & \colhead{0.5$-$8~keV} &\colhead{0.5$-$2~keV} & \colhead{2$-$8~keV} & \colhead{0.5$-$8~keV} & \colhead{0.5$-$2~keV} &\colhead{2$-$8~keV} & \colhead{0.5$-$8~keV} }
\startdata
\multicolumn{10}{c}{\sc \small \PL\ fits} \\
\\
\hline

\irasa & 2.79 & 1.05 & 3.84 & 5.6/3 & 5.7/2 & 5.6/2 & 86.8 & 94.3 & 93.9\\ 
\pga & 1.21 & 10.46 & 11.63 & 2.3/1 & 0.5/1 & 0.7/1 & 86.9 & 50.8 & 58.1\\ 
\pgb & 1.93 & 1.89 & 3.89 & 33.4/2 & 32.8/1 & 32.7/1 & $>$99.9 & $>$99.9 & $>$99.9\\ 
\pgc & 11.27 & 32.50 & 44.74 & 3.0/1 & 87.7/1 & 44.2/1 & 91.5 & $>$99.9 & $>$99.9\\ 
\Qa & 1.54 & 2.76 & 4.38 & 15.7/5 & 12.5/2 & 4.2/2 & 99.2 & 99.8 & 87.5\\ 
\mrk & 11.05 & 42.56 & 53.45 & 18.1/8 & 20.0/8 & 13.1/7 & 98.0 & 99.0 & 93.0\\ 
\irasb & 1.14 & 1.50 & 2.63 & 0.6/2 & 0.6/2 & 0.6/2 & 26.4 & 26.4 & 26.4\\ 
\cso & 8.82 & 12.54 & 21.42 & 5.7/3 & 15.2/4 & 9.6/3 & 87.2 & 99.6 & 97.8\\ 
\SBS & 5.51 & 8.75 & 14.32 & 6.8/3 & 0.1/2 & 0.7/2 & 92.4 & 4.3 & 28.1\\ 
\pge & 0.68 & 0.93 & 1.61 & 2.5/3 & 2.5/3 & 2.5/3 & 53.1 & 52.8 & 52.9\\ 
\pgf & 1.64 & 10.07 & 11.72 & 717.0/8 & 504.9/7 & 560.1/7 & $>$99.9 & $>$99.9 & $>$99.9\\


\hline
\\
\multicolumn{10}{c}{\sc \small  \PL\ fits with cross-calibration uncertainties} \\
\\
\hline

\irasa & 2.63 & 0.98 & 3.62 & 3.4/3 & 3.5/2 & 3.4/2 & 66.5 & 82.5 & 81.9\\ 
\pga & 1.18 & 10.35 & 11.47 & 2.1/1 & 0.4/1 & 0.6/1 & 84.9 & 48.6 & 55.6\\ 
\pgb & 1.61 & 1.56 & 3.20 & 20.0/2 & 19.9/1 & 19.9/1 & $>$99.9 & $>$99.9 & $>$99.9\\ 
\pgc & 11.29 & 31.82 & 44.05 & 0.3/1 & 7.4/1 & 3.8/1 & 38.9 & 99.4 & 94.9\\ 
\Qa & 1.59 & 2.81 & 4.45 & 9.5/5 & 6.8/2 & 2.8/2 & 91.1 & 96.7 & 75.8\\ 
\mrk & 11.11 & 43.58 & 54.39 & 2.5/8 & 3.5/8 & 2.2/7 & 3.9 & 9.9 & 5.2\\ 
\irasb & 1.13 & 1.49 & 2.62 & 0.6/2 & 0.6/2 & 0.6/2 & 25.3 & 25.3 & 25.3\\ 
\cso & 8.24 & 11.52 & 19.92 & 2.1/3 & 7.0/4 & 4.2/3 & 45.7 & 86.4 & 76.4\\ 
\SBS & 5.24 & 8.74 & 14.04 & 3.9/3 & 0.1/2 & 0.3/2 & 72.9 & 2.2 & 14.4\\ 
\pge & 0.67 & 0.92 & 1.60 & 2.2/3 & 2.1/3 & 2.2/3 & 46.2 & 45.9 & 46.0\\ 
\pgf & 1.59 & 9.58 & 11.18 & 146.0/8 & 99.5/7 & 107.1/7 & $>$99.9 & $>$99.9 & $>$99.9\\ 


\hline
\\
\multicolumn{10}{c}{\sc \small \PL\ fits with cross-calibration uncertainties and upper limits} \\
\\
\hline
\irasa & 2.62 & 0.98 & 3.61 & 3.4/4 & 3.5/3 & 3.4/3 & 50.7 & 67.8 & 66.9\\ 
\pga & 1.00 & ... & ... & 3.8/2 & ... & ... & 85.0 & ... & ...\\
\pgb & 1.59 & 1.51 & 3.12 & 21.8/4 & 20.2/2 & 20.1/2 & $>$99.9 & $>$99.9 & $>$99.9\\
\pgc & 10.45 & ... & ... & 9.2/2 & ... & ... & 99.0 & ... & ...\\
\Qa & 1.58 & ... & ... & 11.5/7 & ... & ... & 88.4 & ... & ...\\ 
\mrk & 11.04 & ... & ... & 4.4/9 & ... & ... & 12.0 & ... & ...\\
\irasb & 0.86 & ... & ... & 3.0/3 & ... & ... & 61.4 & ... & ...\\ 
\pge & 0.68 & 0.94 & 1.62 & 3.0/5 & 3.0/4 & 3.0/4 & 30.5 & 44.1 & 44.0\\ 




\hline
\\
\multicolumn{10}{c}{\sc \small fits using complex models with cross-calibration uncertainties} \\
\\
\hline

\mrk & 10.63 & 48.40 & 59.13 & 4.0/8 & 1.2/7 & 1.1/7 & 14.7 & 0.9 & 0.8\\ 
\cso & 8.60 & 11.14 & 19.80 & 1.0/3 & 3.1/3 & 2.0/3 & 20.1 & 63.4 & 42.8\\ 
\pgf & 1.39 & 10.53 & 11.91 & 108.1/8 & 72.6/7 & 81.6/7 & $>$99.9 & $>$99.9 & $>$99.9\\ 
 

\enddata

\tablenotetext{a}{\hbox{Best-fitting} constant flux (weighted average) from the \chandra, \xmm, \sax, \asca, and \rosat\ detections of each source. }

\end{deluxetable*}

\begin{deluxetable*}{cccccccc}
\tablecolumns{7}
\tabletypesize{\scriptsize}
 \tablewidth{0pt}
\tablecaption{Average S/N, Expected Fraction of Variable Sources, and Observed Minimum-to-Maximum Flux Ratios.\label{tab:SNra}}
\tablehead
{
\colhead{\sc object name$^{\rm a}$}& \colhead{$\langle S/N\rangle$} &  \colhead{E.C.$^{\rm b}$} & \colhead{$P_{\rm VD}$$^{\rm c}$} & \colhead{$\left[\frac{f_{\rm max}}{f_{\rm min}}\right]_{0.5-2}$} & \colhead{$\left[\frac{f_{\rm max}}{f_{\rm min}}\right]_{2-8}$} & \colhead{$\left[\frac{f_{\rm max}}{f_{\rm min}}\right]_{0.5-8}$} & 
\vspace{5pt} }

\startdata

\irasa  & 6.6 &  50  & 0.20 & 1.5$\pm$0.4 &  1.5$\pm$0.4 & 1.5$\pm$0.4  \\ 
\pga  &   4.7 &  27 & 0.20 & 2.1$\pm$1.4 & 1.4$\pm$0.7 &  1.4$\pm$0.7   \\ 
\pgb*  & 7.8 & 69 & 0.20  & 3.8$\pm$1.3 & 3.8$\pm$1.3 & 3.8$\pm$1.3 \\ 
\pgc* & 45.1& 2079& 0.80 &  1.1$\pm$0.2 & 1.5$\pm$0.2 & 1.3$\pm$0.2  \\ 
\Qa  & 8.8 & 86  & 0.20  & 2.4$\pm$0.8 &  2.0$\pm$0.5 & 1.6$\pm$0.4 \\ 
\mrk  & 29.8 & 917& 0.43 & 1.2$\pm$0.2 & 1.3$\pm$0.2 & 1.2$\pm$0.2  \\ 
\irasb  & 2.1& 7 & 0.02  & 1.5$\pm$1.0  & 1.5$\pm$1.0 & 1.5$\pm$1.0  \\ 
\cso  & 17.2 & 313 & 0.43 &  1.3$\pm$0.3 & 1.8$\pm$0.4 & 1.5$\pm$0.4 \\ 
\SBS  & 13.4 &  193 & 0.43  & 1.6$\pm$0.5 & 1.1$\pm$0.2 & 1.1$\pm$0.2  \\ 
\pge  & 4.8 & 28 & 0.20  & 1.7$\pm$0.7  & 1.7$\pm$0.7 & 1.7$\pm$0.7 \\ 
\pgf*  & 25.2 & 660 & 0.43 & 17.5$\pm$4.1 & 9.9$\pm$2.3 & 11.1$\pm$2.6  \\ 


\enddata

\tablenotetext{a}{The sources marked with * present evidence of variability in at least one of the energy bands considered.}

\tablenotetext{b}{Expected number of counts in the  \chandra\ full band in order to achieve the average S/N of the set of observations of each source.}
\tablenotetext{c}{Variability detection probabilities obtained from the expected mean number of counts in the \chandra\  full band.  These probabilities are obtained from column 3 of this table and from Tables~1 and 3 of  \cite{2012ApJ...746...54G}.}

\end{deluxetable*}

\begin{deluxetable*}{ccccccc}
\tablecolumns{8}
\tabletypesize{\scriptsize}
 \tablewidth{0pt}
\tablecaption{Derived X-ray Fluxes \label{tab:flcm} from Complex Models Fits}
\tablehead
{
\colhead{\sc object name\tablenotemark{a}} & \colhead{\sc obs. date} & \colhead{$f_{0.5-2}$\tablenotemark{b}} & \colhead{$f_{2-8}$\tablenotemark{b}} & \colhead{$f_{0.5-8}$\tablenotemark{b}} & \colhead{$f_{\rm 2keV}$\tablenotemark{c}} & \colhead{Instrument}
}
\startdata

{\mrk} 
& 1991 Jun 07 & 11.0$\pm$1.6  & ... & ...& 14.8$\pm$3.9 & PSPC \\ 
& 1994 May 12 & 10.8$\pm$1.7 & 46.0$\pm$7.2 & 56.8$\pm$9.0 & 34.7$\pm$5.5  & SIS  \\ 
& 1999 Nov 10 & 13.8$\pm$1.5 & 49.1$\pm$5.4 & 62.9$\pm$6.9 & 23.7$\pm$2.6  & SIS  \\ 
& 2000 Oct 19 & 10.4$\pm$1.1 & 48.9$\pm$5.2 & 59.4$\pm$6.4 & 20.5$\pm$2.2 & ACIS-S  \\
& 2001 Jun 07 & 11.5$\pm$0.5 & 51.9$\pm$2.3 & 63.4$\pm$2.8 & 17.88$\pm$0.79  & EPIC  \\
 & 2003 Feb 03 & 10.22$\pm$0.66 & 44.1$\pm$2.8 & 54.3$\pm$3.5 & 18.3$\pm$1.2 & ACIS-S  \\
  & 2003 Feb 11 & 10.13$\pm$0.60 & 48.8$\pm$2.9 & 58.9$\pm$3.5 & 14.93$\pm$0.89 & ACIS-S  \\
& 2003 Feb 20 & 9.60$\pm$0.66 & 49.6$\pm$3.4 & 59.2$\pm$4.1 & 15.2$\pm$1.1 & ACIS-S  \\
 & 2010 Jul 11 & $10.3_{-1.5}^{+2.1}$ & $49_{-7}^{+10}$ & $59_{-9}^{+12.}$ & $23.5_{-3.5}^{+4.9}$ & ACIS-S      
 \vspace{5pt}   \\
 
{\cso} 
 & 2001 Jul 30 & 8.5$\pm$1.7 & 12.5$\pm$2.6 & 21.0$\pm$4.3 & 26.7$\pm$5.5  & EPIC  \\
 & 2001 Dec 08 & 9.14$\pm$0.43 & 12.46$\pm$0.59 & 21.6$\pm$1.0 & 27.4$\pm$1.3  & EPIC  \\ 
& 2001 Dec 13 & 8.6$\pm$1.2 & 10.4$\pm$1.4 & 19.0$\pm$2.6 & 17.4$\pm$2.4  & EPIC  \\
  & 2010 Mar 21 & 7.1$\pm$1.6 & 8.3$\pm$1.9 & 15.4$\pm$3.5 & 22.3$\pm$5.1 & ACIS-S 
\vspace{5pt}   \\
 
{\pgf*} 
& 1991 Nov 30 & $8.3_{-3.5}^{+5.7}$  & ... & ...& $16_{-9}^{+13}$ & PSPC \\
 & 1999 Oct 30 & 31.3$\pm$4.2 & 63.8$\pm$8.6 & 95$\pm$13 & 75$\pm$10  & SIS  \\ 
& 2002 Sep 01 & 3.06$\pm$0.32 & 14.7$\pm$1.5 & 17.7$\pm$1.8 & 7.14$\pm$0.74 & ACIS-S  \\ 
& 2003 May 14 & 16.3$\pm$1.3 & 34.7$\pm$2.7 & 51.0$\pm$3.9 & 41.2$\pm$3.2  & EPIC  \\
& 2005 Nov 20 & 1.13$\pm$0.08 & 7.54$\pm$0.56 & 8.67$\pm$0.64 & 2.20$\pm$0.16  & EPIC  \\
& 2007 May 03 & 1.62$\pm$0.22 & 12.6$\pm$1.7 & 14.2$\pm$1.9 & 3.67$\pm$0.50  & EPIC  \\
& 2007 May 19 & 1.54$\pm$0.18 & 10.7$\pm$1.2 & 12.3$\pm$1.4 & 4.06$\pm$0.46  & EPIC  \\
& 2007 May 21 & 1.44$\pm$0.11 & 12.38$\pm$0.91 & 13.8$\pm$1.0 & 3.62$\pm$0.26  & EPIC  \\
 & 2007 Nov 05 & 1.12$\pm$0.13 & 9.1$\pm$1.1 & 10.2$\pm$1.2 & 3.02$\pm$0.36  & EPIC

\enddata


\tablecomments{Fluxes are obtained from the best-fitting parameters of  the \hbox{ionized-absorbed} \PL\ model for \cso\  and the thermal plasma + two absorbed \PLs\ model in the case of \mrk\ and \pgf.  
The fluxes are not corrected for intrinsic absorption.}


\tablenotetext{a}{The sources marked with * present evidence 
of variability in at least one of the energy bands considered.}

\tablenotetext{b}{Observed fluxes in the  0.5$-$2, 2$-$8, and 0.5$-$8~keV  bands in units of $10^{-14}$~\flux. }

\tablenotetext{c}{Observed flux densities at \RF\  2~keV in units of $10^{-32}$~\flux Hz$^{-1}$. }

\end{deluxetable*}

\subsection{Long and short-term variability  \label{S:vaPL}}

Long-term variability was assessed initially by comparing the fluxes in the \SB, \HB, and \FB\ bands for all the detections of each source. 
In order to test the significance of any variability, we calculate for each source the $\chi^2$ statistic where the data points are the fluxes of each observation (with their respective 1$\sigma$ errors) and the model is the \hbox{best-fit} constant flux.
The $\chi^2$ value provides a statistical test of the null hypothesis that the flux of each epoch is equal to the best-fit flux (average flux weighted by errors) of all the observations. 

In Table~\ref{tab:fave} we list the \hbox{best-fit} constant fluxes for each source in each band. In this table we also have listed the values of  $\chi^2$ for our model to test variability. 
We find that six out of eleven sources show evidence for variability (significance $>99\%$; see Table~\ref{tab:fave}). The sources with potential variability are \pgb, \pgc, \Qa, \mrk, \cso, and \pgf. Notice that for most of these sources the apparent variability is found between observations performed with different \XR\ instruments (e.g., see Figures~\ref{fig:var1}, \ref{fig:var2}, and \ref{fig:var3}). The only exception is \pgf\ where variability is independently found in the \xmm\ observations  (see Figures~\ref{fig:var1}, \ref{fig:var2}, \ref{fig:var3} and Table~\ref{tab:flu}).  Therefore, in order to assess the variability reliably we must take into account uncertainties associated with cross calibration between the different \XR\ missions used in this work. From studies of different astrophysical sources that can be considered ``standard candles" in the \XR\ regime the main conclusion is that, in general, the \hbox{cross-calibration} errors in the \XR\ fluxes between \chandra, \xmm, \sax, \asca, \rosat, and \einstein\ should be $\lesssim 10\%$ \citep[e.g.,][]{2002astro.ph..3311S,2005SPIE.5898...22K, 2010A&A...523A..22N, 2010ApJ...713..912W, 2011A&A...525A..25T}.  Therefore, as a second test of the variability found we add in quadrature to the error bars of each measured flux a 10\% error. The new constant flux estimates with their $\chi^2$ values and significances are presented in Table~\ref{tab:fave}.   We find that three (\pgb, \pgc, and \pgf) out of the five variable candidates show significant variability (at a level $>99\%$) in the measured fluxes.  

\begin{figure}
   \includegraphics[width=9cm]{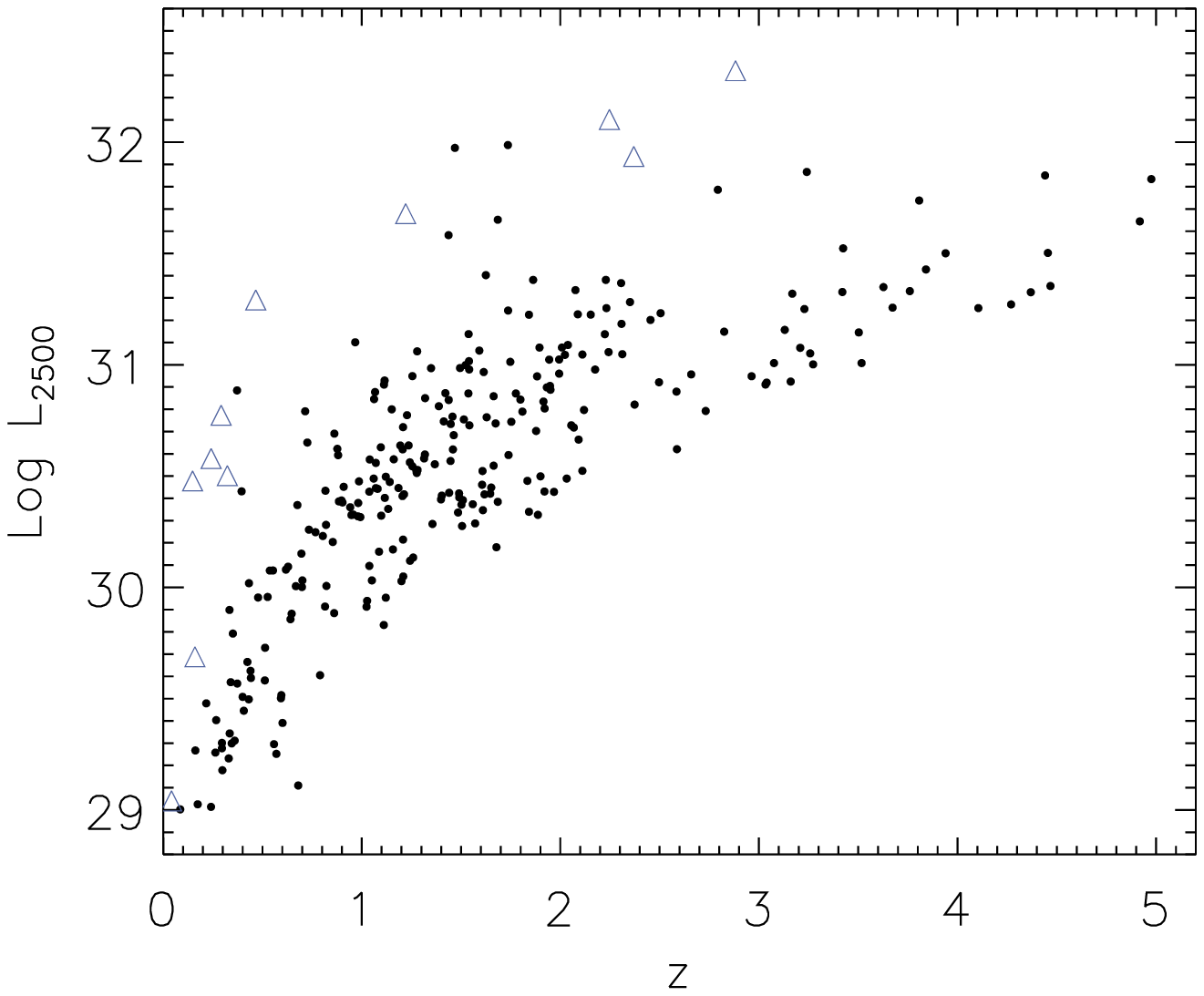}
        \centering
       \caption{Logarithm of the monochromatic luminosity at \RF\  $2500$~\AA\  versus redshift. Monochromatic luminosities used here have units of $\lumin {\rm Hz}^{-1}$. Black-filled circles are data from \cite{2012ApJ...746...54G}.  Our sample is represented with open triangles.}
     \label{fig:Gibs}
     \end{figure}

We have made a basic comparison of the fraction of \XR\ variable BAL quasars in our sample with the larger \chandra\ quasar sample of  \cite{2012ApJ...746...54G}. 
To make an appropriate basic comparison with this sample, we obtain the average S/N of the observations of each source in our sample (see Table~\ref{tab:SNra}). These S/N values are converted to an equivalent number of counts in the \chandra\ full band (see Table~\ref{tab:SNra}).   We assume that the errors in the \chandra\ counts are given by Tables~1 and 2 of \cite{1986ApJ...303..336G}. We proceed by obtaining the expected variability detection probability for each source based on the \hbox{count-dependent} fraction of variable sources found in \S3.2 of \cite{2012ApJ...746...54G}. For this purpose, we have divided the sources in \cite{2012ApJ...746...54G} into four count ranges: \hbox{1--10}, \hbox{10--100}, \hbox{100--1000}, and \hbox{1000--10000} counts. The fractions (variability detection probabilities) of  detectably variable sources  in each range are 0.02, 0.20, 0.43, and 0.80. For each source we select one of these four values based on its equivalent average counts in the \chandra\ full band (see Table~\ref{tab:SNra}).  Adding the probabilities found in Table~\ref{tab:SNra},  the expected number of variable sources is $\approx 3.5$. Therefore, the fraction of \XR\ variable sources appears comparable to that expected for normal quasars. As is shown in \cite{2012ApJ...746...54G}, the fraction of detectably variable sources in \XRs\ has a \hbox{first-order} dependency on the S/N  of the observations of each source. This is because the sensitivity to measure any change in the flux increases with S/N. 
Additionally, the UV luminosities of our sources are broadly comparable to those of the sample of  \cite{2012ApJ...746...54G}, although both sets of sources span large ranges of luminosity (see Figure~\ref{fig:Gibs}). The mean luminosity of our sample ($\langle L_{2500} \rangle=30.95\pm0.31$) is similar to the mean luminosity of the sample of  \cite{2012ApJ...746...54G} ($\langle L_{2500} \rangle=30.55\pm0.04$). Therefore, any secondary luminosity dependency \citep[e.g.,][]{1997ApJ...476...70N, 2002MNRAS.330..390M}\footnote{ Not clear relation was found between luminosity and X-ray variability in  \cite{2012ApJ...746...54G}.} of the variability should not fundamentally affect our basic comparisons with \cite{2012ApJ...746...54G}. 

The ratio of maximum-to-minimum fluxes of the variability is $3.8\pm1.3$, $1.5\pm0.2$, and $9.9\pm2.3$ for \pgb, \pgc, and \pgf, respectively. These ratios have been calculated in the \HB\ band since in this band all the sources present variability. Additionally, the shortest  \RF\ timescales of significant variability were approximately 5.8, 1.4, and 0.5 yr for \pgb, \pgc, and \pgf, respectively. Since in general there is not extensive coverage of sources with a single \XR\ mission (with the exception of \pgf\ and \mrk), the flux uncertainties of our variability analysis are greater than 10\%.
Hence, for most of the sources we are not sensitive to amplitude changes $\lesssim 1.4$.\footnote{Assuming two flux measurements with a fractional error of 10\% each, our variability test requires  a  \hbox{maximum-to-minimum} flux ratio of $\gtrsim1.4$ in order to get a $\gtrsim99\%$ level of significance.}  In general the amplitude threshold for variability detection is higher than 1.4, especially for observations with less than $\sim 100$~counts, for which errors in the flux are above 10\%. As found in Table~\ref{tab:SNra} the \hbox{maximum-to-minimum} flux amplitudes of the sources that do not present statistically significant variability are 10--100\% in the \HB~band.

Notice that in the variability searching above we have not taken into account  flux upper limits. In general, as seen from Figures \ref{fig:var1}, \ref{fig:var2}, and \ref{fig:var3}, the typically loose flux upper limits do not present any new useful information regarding variability; specifically, the upper limits are usually above or not far below the best-fit constant flux.  As one basic illustrative test to see if the upper limits give any new information about variability, we assume that the upper limits  are detections with fluxes given by one half of the threshold  and error bars that are three tenths of the \hbox{upper-limit} value.\footnote{If we assume that the flux below the \hbox{upper-limit} has a uniform distribution, the mean value is half the threshold and $\sigma$ is $\sqrt{1/12} \approx 0.28$ of the \hbox{upper-limit}.} Under this assumption we test for variability by calculating the fits of constant flux in Table~\ref{tab:fave}. The test with the upper limits presented in Table~\ref{tab:fave} is with 10\% cross-calibration errors taken into account. From this test we find that the upper limits are possibly giving new information about variability in just one case.  This case corresponds to the \einstein\ upper limit for \pgc. From Figure \ref{fig:var1} and Table~\ref{tab:fave}  it is clear that the upper limit increases the significance of the soft-band variability of \pgc. Since this source has already shown variability in the hard band and possibly full band, adding the upper limit does not contribute any material change to the conclusions of the variability analysis.

We also searched for rapid variability in the high \hbox{signal-to-noise} ratio ($\rm S/N > 25$) \chandra, \xmm, \asca, and \rosat\  observations by applying the \hbox{Kolmogorov-Smirnov} test to the photon arrival times. No significant variability (at a level of significance $>$99\%) in the full band of each instrument was found.

As illustrated with the shaded regions in Figures \ref{fig:var1}, \ref{fig:var2}, and \ref{fig:var3},
the BAL quasars in our sample are usually X-ray weak by factors
of \hbox{$\approx 2$--100} relative to expectations for
non-BAL quasars (see \S\ref{S:vaan} for further discussion). Thus, any
substantial removal of the shielding gas responsible for this
X-ray weakness (see \S\ref{S:intr}) from our line of sight should lead
to large X-ray flux changes---changes that are larger
than the upper limits that we typically place above upon
allowed X-ray variability. The fact that such large X-ray flux
changes appear rare over \hbox{3--30~yr} suggests that the
shielding gas is fairly stable, even on timescales sufficiently
long to allow significant physical rearrangement and
accretion-disk rotation. This fact should be considered in
future modeling of BAL quasar X-ray absorption variability
\citep[e.g.,][]{2010MNRAS.408.1396S} as well as in interpretation of
UV BAL variability results \citep[e.g.,][]{2008ApJ...675..985G, 2010ApJ...713..220G, 2011MNRAS.413..908C, 2012MNRAS.422.3249C, 2012arXiv1208.0836F}.

\begin{figure*}
   \includegraphics[width=16cm]{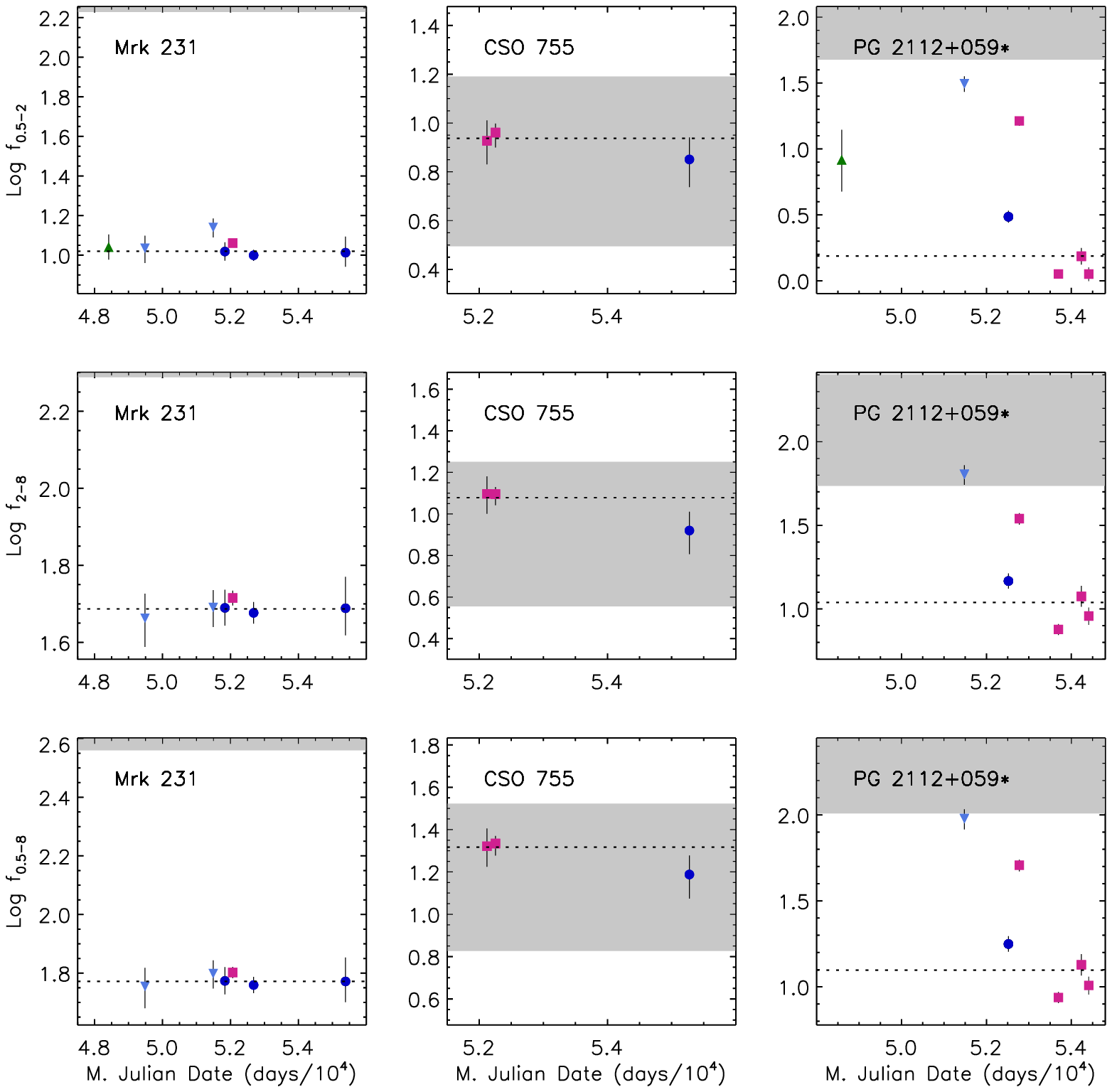}
        \centering
    \caption{Logarithm of the observed flux  (in units of $10^{-14}$~\flux) versus modified Julian date for \mrk, \cso, and \pgf.  The fluxes were obtained using alternative models to a simple \PL. The three upper, middle, and lower panels correspond to fluxes in the soft (\SB), hard (\HB) and full (\FB) bands, respectively. The circles, squares, inverted triangles, and triangles indicate observations performed with \chandra, \xmm, \asca, and \rosat, respectively.  See  the Figure~\ref{fig:var1} caption for additional details about the presentation.}
     \label{fig:var4}
     \end{figure*}

\subsection{Flux variability in complex spectra\label{S:vaCM}}

For low S/N observations it is not possible to probe effectively spectral complexity that deviates from the \PL\  models used in \S\ref{S:span}. However, in cases where the S/N is sufficiently high we can quantify these deviations. We address this issue by grouping with a minimum of ten counts per bin the high S/N observations ($\rm S/N > 25$) and then performing $F$-tests to assess the improvement of the fits from a \PL\ through the use of more complex alternative models. 

 \begin{deluxetable*}{ccccccccccccc}
\tablecolumns{13}
\tabletypesize{\scriptsize}
 \tablewidth{0pt}
\tablecaption{Optical and \XR\ Properties of Surveyed BAL Quasars  \label{tab:opxr}}
\tablehead
{
\colhead{\sc object name} & \colhead{Log~$L_{\rm 2500}$\tablenotemark{a}} & \colhead{Log~$L_{\rm 2-10}$\tablenotemark{b}} & \colhead{Log~$L_{\rm 2 keV}$\tablenotemark{c}} & \colhead{$\Delta$Log~$L_{\rm 2 keV}$\tablenotemark{d}} & \colhead{Log~$f_{2-8}$\tablenotemark{e}} &  \colhead{$\Delta$Log~$f_{2-8}$\tablenotemark{f}}  & \colhead{\aox \tablenotemark{g}}& \colhead{$\Delta \aox$\tablenotemark{h}} &
}
\startdata

\irasa  & 30.48 & 41.87 & 24.18 & $-2.26$ & $-14.01$ & $-2.53$ & $-2.42$$\pm$0.06 &  $-0.85$ \\ 
\pga  & 31.68 & 44.74 & 25.42 & $-1.86$ & $-12.99$ & $-0.23$ & $-2.40$$\pm$0.08 &  $-0.66$ \\ 
\pgb  & 29.69 & 42.34 & 24.22 & $-1.65$ & $-13.81$ & $-1.70$ & $-2.10$$\pm$0.07 &  $-0.64$ \\ 
\pgc  & 30.58 & 43.82 & 25.61 & $-0.89$ & $-12.50$ & $-0.62$ & $-1.91$$\pm$0.05 &  $-0.32$ \\ 
\Qa  & 32.10 & 44.94 & 26.65 & $-0.93$ & $-13.55$ & $-0.46$ & $-2.09$$\pm$0.05 &  $-0.29$ \\ 
\mrk  & 29.04 & 42.36 & 24.08 & $-1.33$ & $-12.36$ & $-1.02$ & $-1.90$$\pm$0.04 &  $-0.53$ \\ 
\irasb  & 30.50 & 42.78 & 24.80 & $-1.65$ & $-13.83$ & $-1.61$ & $-2.19$$\pm$0.10 &  $-0.62$ \\ 
\cso  & 32.32 & 45.88 & 27.63 & $-0.11$ & $-12.94$ & $0.25$ & $-1.80$$\pm$0.05 &  $0.02$ \\ 
\SBS  & 31.94 & 45.51 & 27.32 & $-0.15$ & $-13.06$ & $0.21$ & $-1.77$$\pm$0.05 &  $-0.00$ \\ 
\pge  & 30.77 & 42.48 & 24.48 & $-2.16$ & $-14.03$ & $-2.11$ & $-2.42$$\pm$0.06 &  $-0.80$ \\ 
\pgf  & 31.29 & 43.99 & 25.40 & $-1.61$ & $-13.02$ & $-0.99$ & $-2.26$$\pm$0.04 &  $-0.57$ \\


\enddata

\tablenotetext{a}{\hbox{Logarithm} of the monochromatic luminosity at \RF\ 2500 \AA\ (with units \lumin~Hz$^{-1}$).  These values were computed from the  flux densities at \RF\ wavelength 2500 \AA\  using the corrected for galactic reddening flux densities of the closest optical magnitude and  extrapolating a \PL\ with $\alpha=-0.5$ \citep[e.g.,][]{2001AJ....122..549V}. }

\tablenotetext{b}{\hbox{Logarithm} of the 2$-$10~keV band luminosity (with units \lumin). For each source this quantity was obtained from the weighted-average luminosities of the \chandra, \xmm, and \asca\ observations in the 2$-$10~keV \RF\ band. The luminosities have been corrected for Galactic absorption. } 

\tablenotetext{c}{\hbox{Logarithm} of the monochromatic luminosity at \RF\ 2~keV (with units \lumin~Hz$^{-1}$). For each source this quantity was obtained from the \hbox{weighted-average} flux of the \chandra, \xmm, \asca, and \rosat\ observations at 2~keV \RF.}

\tablenotetext{d}{The difference between the measured and the predicted log~$L_{\rm 2  keV}$ for RQQs, based on the \cite{2007ApJ...665.1004J} relation: \hbox{${\rm log}~L_{\rm 2  keV}=0.709 \times {\rm log}~L_{\rm 2500}+4.822$} .}

\tablenotetext{e}{\hbox{Logarithm} of the flux in the 2$-$8~keV band (in units of $10^{-14}$~\flux). For each source this flux was obtained from the weighted-average flux of the \chandra, \xmm, \sax, and \asca\ observations in the 2$-$8~keV band.} 

\tablenotetext{f}{The difference between the measured and the predicted flux in the 2$-$8~keV band for RQQs. The predicted fluxes in the 2$-$8~keV band are obtained from the expected fluxes for RQQs at  2~keV \RF\  \citep{2007ApJ...665.1004J} assuming a \PL\ spectrum with $\Gamma=1.9$.}

\tablenotetext{g}{The optical-to-X-ray power-law slope $\aox= {\rm log}(f_{\rm 2keV}/f_{\hbox{\tiny 2500\AA}})/{\rm log}(\nu_{\rm 2keV}/\nu_{\hbox{\tiny 2500\AA}})$.}

\tablenotetext{h}{The difference between the measured \aox\ value and the predicted \aox\ for RQQ, based on the \cite{2007ApJ...665.1004J} relation:  \mbox{$\aox=-0.140 \times {\rm log}~L_{\hbox{\tiny 2500\AA}}+2.705$}.}

\end{deluxetable*}

The alternative models  selected are used to probe complex absorption and to corroborate previous studies of sources in our sample \citep[e.g.,][]{2000ApJ...545L..23M, 2002ApJ...569..655G, 2004ApJ...603..425G, 2005A&A...433..455S, 2010A&A...512A..75S, 2006ApJ...652..163M}.\footnote{\label{fo:amod} These models are (1) absorbed \PL\ with iron emission line, (2) ionized-absorbed \PL\ (\xspec\ model {\sc absori}), (3) partially covered absorbed \PL, (4) reflection spectrum (\xspec\ model {\sc pexrav}), (5) thermal plasma (\xspec\ model {\sc vraymond}) + \PL, (6) thermal plasma + two absorbed \PLs\ (scattered spectra),  (7)  thermal plasma + reflection spectrum, (8) reflection spectrum from ionized media (\xspec\ model {\sc pexriv}), and (9) \PL\ with absorption from a blueshifted iron line.} Through this method we find that in three sources at least one of the alternative models improves the fit with significance \mbox{$>99\%$} in every high S/N observation; these sources are \mrk, \cso, and \pgf.   In the case of \mrk\ there were two models (models 6 and 7 in Footnote~\ref{fo:amod}) that provide improvement of the fits in every high S/N observation. These models are thermal plasma + two absorbed \PLs\  and  thermal plasma + reflection spectrum. Note that when we use a model with two absorbed \PLs,  each \PL\ component has the photon index fixed to \mbox{$\Gamma=2$} and the column density free to vary.  The \PL\ that is heavily absorbed represents the intrinsic X-ray emission and the second \PL\  with less absorption represents scattered emission \citep[see, e.g.,][]{2002ApJ...567...37G, 2004A&A...420...79B}.  In the case of \cso\  an ionized absorbed \PL\  model improves the fit in the only high S/N observation \citep[in agreement with][]{2005AJ....130.2522S}.
In the case of \pgf\ there were two models that show improvement of the fits in every high S/N observation.  These models are thermal plasma + two absorbed \PLs\  and reflection spectrum from ionized media. 

Based on these fitting results, we have recalculated the fluxes in each observation and band for \mrk\  and \pgf\ using the thermal plasma + two absorbed \PL\ model,  and for \cso\ using an \hbox{ionized-absorbed} \PL\ model.\footnote{The error in the estimated fluxes using the thermal plasma + two \PL\ model depends mainly on the normalization of the two \PLs\ ($n_{\rm PL1}$ and $n_{\rm PL2}$), and therefore we estimate the error on the flux in a given band ($f_{\rm BP}$) by $\Delta f_{\rm BP}=|\frac{\partial f_{\rm BP}}{\partial n_{\rm PL1}}|\Delta n_{\rm PL1}+|\frac{\partial f_{\rm BP}}{\partial n_{\rm PL2}}|\Delta n_{\rm PL2}$.}
The fluxes obtained using these new models are presented in Table~\ref{tab:flcm} and plotted in Figure~\ref{fig:var4}. The models used to obtain the fluxes in  Figure~\ref{fig:var4} (and Table~\ref{tab:flcm}) involve more parameters than a simple \PL; therefore, the error bars on the fluxes in this figure are in general somewhat larger than those obtained in Figures~\ref{fig:var1}, ~\ref{fig:var2}, and ~\ref{fig:var3}.  The \hbox{flux-variability} analysis for \mrk, \cso, and \pgf\ using complex models  is consistent  with that using a simple \PL\ model (see Table~\ref{tab:fave} and \S \ref{S:vaPL}). Additionally,  the weighted average fluxes obtained for \mrk, \cso, and \pgf\ through the use of more complex models are within 20\% of those obtained using a \PL\ model (see Table~\ref{tab:fave}). Therefore, as expected, our main variability results appear robust to alternative modeling of the spectra.

\begin{figure*}
   \includegraphics[width=15cm]{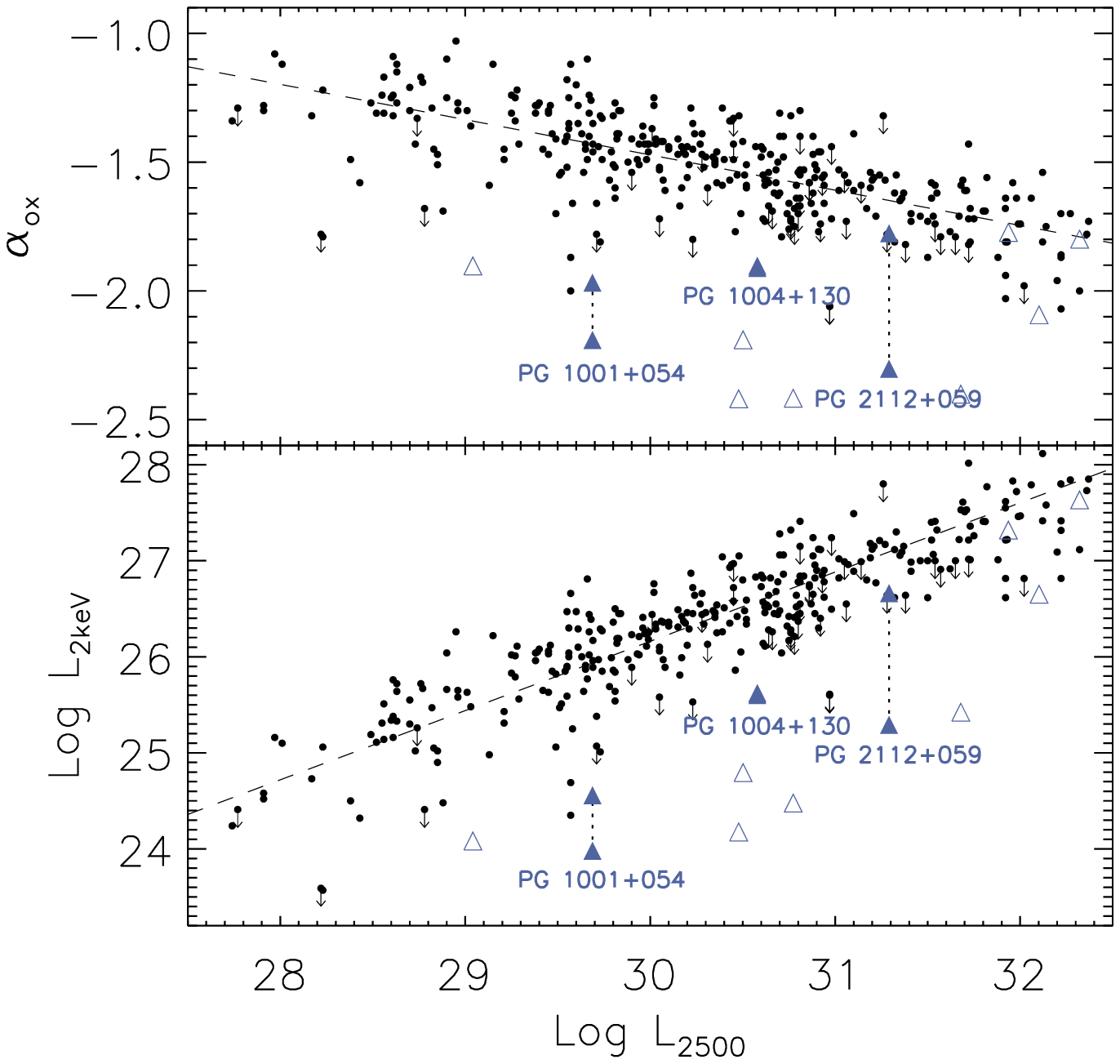}
        \centering
       \caption{Logarithm of the monochromatic luminosity at \RF\  $2500$~\AA\  versus \aox\ (upper panel) and logarithm of the monochromatic luminosity at \RF\ 2~keV (lower panel). Monochromatic luminosities used here have units of \lumin~Hz$^{-1}$. Black-filled circles are data from \cite{2006AJ....131.2826S}; arrows indicate \XR\ upper limits. Dashed lines are the best-fit linear relations for their combined sample. Our sample is represented with filled and open triangles for sources with and without variability, respectively. For variable sources we have plotted the state of minimum and maximum flux joined with a dotted line. \pgc\ does not show variability in soft \XRs, and therefore its  2~keV luminosity is presented as a single point.}
     \label{fig:Stef}
     \end{figure*}

\subsection{Spectral variability\label{S:vaan}}

As expected our sample of BAL quasars is generally  X-ray weak when  compared with RQQs of similar UV luminosities (see Table~\ref{tab:opxr} and Figures~\ref{fig:var1}, \ref{fig:var2}, \ref{fig:var3}, and \ref{fig:Stef}).  On average our sample has luminosities at \RF\ 2~keV that are $\approx 20$ times weaker  than expectations for RQQs \citep{2006AJ....131.2826S, 2007ApJ...665.1004J}. With the exceptions of \cso\ and \SBS, which are as X-ray luminous as typical RQQs, all of our sample is $\gtrsim 3\sigma$ below the \hbox{$\aox-L_{\rm UV}$} and \hbox{$L_{\rm 2keV}-L_{\rm UV}$} relations  \citep[see Table~\ref{tab:opxr} and Table 5 of][]{2006AJ....131.2826S}.  In Figure~\ref{fig:Stef} we have plotted  \aox\ and $L_{\rm 2keV}$ versus $L_{UV}$ for our sample. For the cases of X-ray variable sources we have marked the states of maximum and minimum \XR\ flux in the \hbox{$\aox-L_{\rm UV}$} and \hbox{$L_{\rm 2keV}-L_{\rm UV}$} diagrams assuming that the UV luminosity is constant. The assumption that the UV luminosity is constant is required given that the \XR\ observations do not have simultaneous optical/UV follow-up in general. However,  in general it is expected that the luminosity in the UV  will vary much less than that in the X-rays. For example, the \xmm\ observations of \pgf\ present \hbox{maximum-to-minimum} flux ratios in the X-rays of $\approx 6$  and in the optical of $\approx 1.1$ \citep[][]{2010A&A...512A..75S}.

To perform a basic test for \XR\ spectral variability, we utilize the effective photon index from a simple \PL\ fit, recognizing that this photon index may not have a direct physical interpretation in cases of complex spectra.
Using the same variability test as in \S\ref{S:vaPL}, we conclude that changes in $\Gamma$ have significances $>99\%$ for \pgc\ and \pgf.\footnote{We have taken into account the \hbox{cross-calibration} errors in $\Gamma$ which are estimated to be $\lesssim 5\%$ \citep[e.g.,][]{2002astro.ph..3311S, 2005SPIE.5898...22K, 2011A&A...525A..25T}. } For \pgf\ the changes in $\Gamma$ are  confirmed with significant variability ($>99\%$ significance) in the complex fits \hbox{hard-to-soft} flux ratios ($f_{2-8}/f_{0.5-2}$) (see, e.g., Table~\ref{tab:flcm}). In the case of \pgb, the low S/N of the \chandra\ observation does not allow us to constrain the photon index well, and therefore we cannot tell if there is  any change in the shape of the spectra. 
We conclude that the flux variability of \pgc\ and \pgf\ shows associated spectral variability.

From our spectral analyses alone it is not possible to explain the extreme weakness of the X-ray luminosities of our sources; however, this is likely associated with complex absorption. As already noted in \S \ref{S:span}  almost half of our sample (five out of eleven) presents direct evidence of absorption. Additionally, the flat effective spectral indices measured for most of the sample could be an indication of absorption. For example, as described in \cite{2006ApJ...652..163M}, the hard spectrum of \pgc\ could imply partially covering absorption. Furthermore, based on our spectral analyses it is likely that the observations of \mrk\ and the weak states of \pgf\ correspond to epochs of heavy absorption ($\nh > 10^{23}$~\cmsq).  Therefore, the complexity in the observed spectra of these sources could be the result  of heavily absorbed emission plus scattered emission (see \S \ref{S:vaCM}). In the case of \mrk\ this picture has been shown to be plausible in the joint spectral analysis of \xmm\ and \sax\ observations at energies between \hbox{0.2--50~keV} performed by \cite{2004A&A...420...79B}.  The spectral softening of \pgf\ with increasing flux might be explained by an increase in the observable \XR\ emission due to a decrease in the absorption (see \S \ref{S:vaCM}). Alternative interpretations  of the physical picture for \pgf\ (e.g., a reflection dominated spectrum) can be found in \cite{2010A&A...512A..75S}.  
The \asca\ observation of \pgf\ presents a typical quasar spectral shape with absorption that is moderate enough so the majority of the \XR\ emission can get through the absorber.  Therefore, the \XR\ emission of \pgf\  in its brightest state is likely to be dominated by unabsorbed emission over scattered emission.  The hardening of the spectra with increasing flux for \pgc\  is not easily explained with absorption alone; its origin could be more complex due to \XR\ \hbox{jet-linked} emission \citep[for details see][]{2006ApJ...652..163M}. 
Future \XR\ observations, especially with high sensitivity above 10~keV  such as those with \nustar\ and \astroh, will help to clarify the physical picture of these sources.

\section{SUMMARY AND FUTURE WORK} \label{S:conc}

We have analyzed \chandra, \xmm, \sax, \asca, \rosat, and 
\einstein\ observations of eleven BAL quasars to perform the
first systematic investigation of the long-term \XR\ variability 
of this population on rest-frame timescales of \hbox{3--30~yr}. 
For seven out of the eleven sources we have also obtained and
reported on new \chandra\ observations suitable for detecting 
any strong \XR\ variability. Our main conclusions are the 
following: 

\begin{enumerate}

\item   
We find that three (\pgb, \pgc, and \pgf) of the eleven sources 
show significant flux variability (at a confidence level $>99$\%). 
The maximum observed amplitude of this variability 
in the \HB\ band is a factor of $3.8\pm1.3$, $1.5\pm0.2$, and $9.9\pm2.3$ 
for \pgb, \pgc, and \pgf, respectively. These
three sources show detectable variability on rest-frame timescales
down to 5.8, 1.4, and 0.5~yr, respectively. See \S4.1 and \S4.2. 

\item 
The flux variability of \pgc\ and \pgf\ has associated spectral 
variability (the data for \pgb\ do not allow effective investigation
of spectral variability). This spectral variability may be induced by 
absorption changes, although the hardening of \pgc\ when it brightens
is puzzling in this regard. See \S4.3.

\item
The eight sources in our sample without significantly detected
\XR\ flux variability are constrained by our data to have 
varied by less than \hbox{10--100}\%. See \S4.1.

\item 
We do not find that BAL quasars exhibit exceptional long-term \XR\ 
variability when comparing the measured variability for our whole
sample to that of non-BAL quasars. Strong changes in the \XR\ absorbing 
shielding gas, owing to physical rearrangement or accretion-disk 
rotation, appear relatively rare (although some changes are present). 
Since the shielding gas is critical for setting the ionization level 
of the wind that produces UV BALs, our results have implications for 
modeling BAL variability. See \S4.1. 

\item 
We report for the first time an \XR\ detection of the highly polarized
and well-studied BAL quasar \irasb. We detect this source with a 
statistical significance of $\approx 5 \sigma$ in our new \chandra\ 
observation. See \S2.6. 
 
\end{enumerate}

This first systematic investigation has significantly improved
constraints upon the long-term \XR\ variability of BAL quasars in general. 
However, further improvements are needed in the number of objects 
monitored, the number of monitoring epochs, the number of detected 
counts per epoch, and the bandpass coverage. Improving the numbers 
of objects and epochs will set better constraints upon the frequency 
of significant absorption changes; this could be achieved with, e.g., 
additional \chandra\ and \xmm\ observations. Improving the number 
of detected counts per epoch or extending the the bandpass coverage 
to higher energies will enable better physical understanding when 
absorption changes are detected; missions including \nustar, \astroh, 
and \athena\ are required for such observations.

\acknowledgments
We thank the referee for constructive feedback. We gratefully acknowledge support from NASA ADP grant NNX10AC99G (CS, WNB), NSF grant AST-1108604 (WNB), and NASA grant SAO SV4-74018 (GPG, Principal Investigator). SCG thanks the Natural Science and Engineering Research Council of Canada and the Ontario Early Researcher Award Program for their support. FEB and CS acknowledge support from Programa de Financiamiento Basal and CONICYT-Chile under grants FONDECYT 1101024 and 3120198 and FONDAP-CATA 15010003.
 This research has made use of the Tartarus (Version 3.1) database, created by Paul O'Neill and Kirpal Nandra at Imperial College London, and Jane Turner at NASA/GSFC. Tartarus was supported by funding from PPARC, and NASA grants NAG5-7385 and NAG5-7067.

\clearpage


\begin{thebibliography}{76}
\expandafter\ifx\csname natexlab\endcsname\relax\def\natexlab#1{#1}\fi

\bibitem[{{Avni}(1976)}]{1976ApJ...210..642A}
{Avni}, Y. 1976, \apj, 210, 642

\bibitem[{{Ballo} {et~al.}(2011){Ballo}, {Piconcelli}, {Vignali}, \&
  {Schartel}}]{2011MNRAS.415.2600B}
{Ballo}, L., {Piconcelli}, E., {Vignali}, C., \& {Schartel}, N. 2011, \mnras,
  415, 2600

\bibitem[{{Barlow}(2004)}]{2004physics...6120B}
{Barlow}, R. 2004, ArXiv Physics e-prints/0406120v1

\bibitem[{{Boksenberg} {et~al.}(1977){Boksenberg}, {Carswell}, {Allen},
  {Fosbury}, {Penston}, \& {Sargent}}]{1977MNRAS.178..451B}
{Boksenberg}, A., {Carswell}, R.~F., {Allen}, D.~A., {Fosbury}, R.~A.~E.,
  {Penston}, M.~V., \& {Sargent}, W.~L.~W. 1977, \mnras, 178, 451

\bibitem[{{Boroson} \& {Meyers}(1992)}]{1992ApJ...397..442B}
{Boroson}, T.~A., \& {Meyers}, K.~A. 1992, \apj, 397, 442

\bibitem[{{Braito} {et~al.}(2004){Braito}, {Della Ceca}, {Piconcelli},
  {Severgnini}, {Bassani}, {Cappi}, {Franceschini}, {Iwasawa}, {Malaguti},
  {Marziani}, {Palumbo}, {Persic}, {Risaliti}, \&
  {Salvati}}]{2004A&A...420...79B}
{Braito}, V., {et~al.} 2004, \aap, 420, 79

\bibitem[{{Brandt} {et~al.}(1999){Brandt}, {Comastri}, {Gallagher}, {Sambruna},
  {Boller}, \& {Laor}}]{1999ApJ...525L..69B}
{Brandt}, W.~N., {Comastri}, A., {Gallagher}, S.~C., {Sambruna}, R.~M.,
  {Boller}, T., \& {Laor}, A. 1999, \apjl, 525, L69

\bibitem[{{Broos} {et~al.}(2010){Broos}, {Townsley}, {Feigelson}, {Getman},
  {Bauer}, \& {Garmire}}]{2010ApJ...714.1582B}
{Broos}, P.~S., {Townsley}, L.~K., {Feigelson}, E.~D., {Getman}, K.~V.,
  {Bauer}, F.~E., \& {Garmire}, G.~P. 2010, \apj, 714, 1582

\bibitem[{{Capellupo} {et~al.}(2011){Capellupo}, {Hamann}, {Shields},
  {Rodr{\'{\i}}guez Hidalgo}, \& {Barlow}}]{2011MNRAS.413..908C}
{Capellupo}, D.~M., {Hamann}, F., {Shields}, J.~C., {Rodr{\'{\i}}guez Hidalgo},
  P., \& {Barlow}, T.~A. 2011, \mnras, 413, 908

\bibitem[{{Capellupo} {et~al.}(2012){Capellupo}, {Hamann}, {Shields},
  {Rodr{\'{\i}}guez Hidalgo}, \& {Barlow}}]{2012MNRAS.422.3249C}
---. 2012, \mnras, 422, 3249

\bibitem[{{Cash}(1979)}]{1979ApJ...228..939C}
{Cash}, W. 1979, \apj, 228, 939

\bibitem[{{Chartas} {et~al.}(2009){Chartas}, {Saez}, {Brandt}, {Giustini}, \&
  {Garmire}}]{2009ApJ...706..644C}
{Chartas}, G., {Saez}, C., {Brandt}, W.~N., {Giustini}, M., \& {Garmire}, G.~P.
  2009, \apj, 706, 644

\bibitem[{{Condon} {et~al.}(1998){Condon}, {Cotton}, {Greisen}, {Yin},
  {Perley}, {Taylor}, \& {Broderick}}]{1998AJ....115.1693C}
{Condon}, J.~J., {Cotton}, W.~D., {Greisen}, E.~W., {Yin}, Q.~F., {Perley},
  R.~A., {Taylor}, G.~B., \& {Broderick}, J.~J. 1998, \aj, 115, 1693

\bibitem[{{Di Matteo} {et~al.}(2005){Di Matteo}, {Springel}, \&
  {Hernquist}}]{2005Natur.433..604D}
{Di Matteo}, T., {Springel}, V., \& {Hernquist}, L. 2005, \nat, 433, 604

\bibitem[{{Dickey} \& {Lockman}(1990)}]{1990ARA&A..28..215D}
{Dickey}, J.~M., \& {Lockman}, F.~J. 1990, \araa, 28, 215

\bibitem[{{Elvis} \& {Fabbiano}(1984)}]{1984ApJ...280...91E}
{Elvis}, M., \& {Fabbiano}, G. 1984, \apj, 280, 91

\bibitem[{{Everett}(2005)}]{2005ApJ...631..689E}
{Everett}, J.~E. 2005, \apj, 631, 689

\bibitem[{{Fan} {et~al.}(2009){Fan}, {Wang}, {Wang}, {Wang}, {Dong}, {Zhang},
  \& {Cheng}}]{2009ApJ...690.1006F}
{Fan}, L.~L., {Wang}, H.~Y., {Wang}, T., {Wang}, J., {Dong}, X., {Zhang}, K.,
  \& {Cheng}, F. 2009, \apj, 690, 1006

\bibitem[{{Filiz Ak} {et~al.}(2012){Filiz Ak}, {Brandt}, {Hall}, {Schneider},
  {Anderson}, {Gibson}, {Lundgren}, {Myers}, {Petitjean}, {Ross}, {Shen},
  {York}, {Bizyaev}, {Brinkmann}, {Malanushenko}, {Oravetz}, {Pan}, {Simmons},
  \& {Weaver}}]{2012arXiv1208.0836F}
{Filiz Ak}, N., {et~al.} 2012, ApJ, in press

\bibitem[{{Gallagher}\noopsort{Gallaghera}
  {et~al.}(1999){Gallagher}\noopsort{Gallaghera}, {Brandt}, {Sambruna},
  {Mathur}, \& {Yamasaki}}]{1999ApJ...519..549G}
{Gallagher}\noopsort{Gallaghera}, S.~C., {Brandt}, W.~N., {Sambruna}, R.~M.,
  {Mathur}, S., \& {Yamasaki}, N. 1999, \apj, 519, 549

\bibitem[{{Gallagher}\noopsort{Gallagherb}
  {et~al.}(2001){Gallagher}\noopsort{Gallagherb}, {Brandt}, {Laor}, {Elvis},
  {Mathur}, {Wills}, \& {Iyomoto}}]{2001ApJ...546..795G}
{Gallagher}\noopsort{Gallagherb}, S.~C., {Brandt}, W.~N., {Laor}, A., {Elvis},
  M., {Mathur}, S., {Wills}, B.~J., \& {Iyomoto}, N. 2001, \apj, 546, 795

\bibitem[{{Gallagher}\noopsort{Gallagherc}
  {et~al.}(2002){Gallagher}\noopsort{Gallagherc}, {Brandt}, {Chartas}, \&
  {Garmire}}]{2002ApJ...567...37G}
{Gallagher}\noopsort{Gallagherc}, S.~C., {Brandt}, W.~N., {Chartas}, G., \&
  {Garmire}, G.~P. 2002, \apj, 567, 37

\bibitem[{{Gallagher}\noopsort{Gallagherd}
  {et~al.}(2002){Gallagher}\noopsort{Gallagherd}, {Brandt}, {Chartas},
  {Garmire}, \& {Sambruna}}]{2002ApJ...569..655G}
{Gallagher}\noopsort{Gallagherd}, S.~C., {Brandt}, W.~N., {Chartas}, G.,
  {Garmire}, G.~P., \& {Sambruna}, R.~M. 2002, \apj, 569, 655

\bibitem[{{Gallagher}\noopsort{Gallaghere}
  {et~al.}(2004){Gallagher}\noopsort{Gallaghere}, {Brandt}, {Wills},
  {Charlton}, {Chartas}, \& {Laor}}]{2004ApJ...603..425G}
{Gallagher}\noopsort{Gallaghere}, S.~C., {Brandt}, W.~N., {Wills}, B.~J.,
  {Charlton}, J.~C., {Chartas}, G., \& {Laor}, A. 2004, \apj, 603, 425

\bibitem[{{Gallagher}\noopsort{Gallagherf}
  {et~al.}(2005){Gallagher}\noopsort{Gallagherf}, {Schmidt}, {Smith}, {Brandt},
  {Chartas}, {Hylton}, {Hines}, \& {Brotherton}}]{2005ApJ...633...71G}
{Gallagher}\noopsort{Gallagherf}, S.~C., {Schmidt}, G.~D., {Smith}, P.~S.,
  {Brandt}, W.~N., {Chartas}, G., {Hylton}, S., {Hines}, D.~C., \&
  {Brotherton}, M.~S. 2005, \apj, 633, 71

\bibitem[{{Gallagher}\noopsort{Gallagherg}
  {et~al.}(2006){Gallagher}\noopsort{Gallagherg}, {Brandt}, {Chartas},
  {Priddey}, {Garmire}, \& {Sambruna}}]{2006ApJ...644..709G}
{Gallagher}\noopsort{Gallagherg}, S.~C., {Brandt}, W.~N., {Chartas}, G.,
  {Priddey}, R., {Garmire}, G.~P., \& {Sambruna}, R.~M. 2006, \apj, 644, 709

\bibitem[{{Ganguly} \& {Brotherton}(2008)}]{2008ApJ...672..102G}
{Ganguly}, R., \& {Brotherton}, M.~S. 2008, \apj, 672, 102

\bibitem[{{Gehrels}(1986)}]{1986ApJ...303..336G}
{Gehrels}, N. 1986, \apj, 303, 336

\bibitem[{{Gibson}\noopsort{Gibsona} {et~al.}(2008){Gibson}\noopsort{Gibsona},
  {Brandt}, {Schneider}, \& {Gallagher}}]{2008ApJ...675..985G}
{Gibson}\noopsort{Gibsona}, R.~R., {Brandt}, W.~N., {Schneider}, D.~P., \&
  {Gallagher}, S.~C. 2008, \apj, 675, 985

\bibitem[{{Gibson}\noopsort{Gibsonb} {et~al.}(2009){Gibson}\noopsort{Gibsonb},
  {Jiang}, {Brandt}, {Hall}, {Shen}, {Wu}, {Anderson}, {Schneider}, {Vanden
  Berk}, {Gallagher}, {Fan}, \& {York}}]{2009ApJ...692..758G}
{Gibson}\noopsort{Gibsonb}, R.~R., {et~al.} 2009, \apj, 692, 758

\bibitem[{{Gibson}\noopsort{Gibsond} {et~al.}(2010){Gibson}\noopsort{Gibsond},
  {Brandt}, {Gallagher}, {Hewett}, \& {Schneider}}]{2010ApJ...713..220G}
{Gibson}\noopsort{Gibsond}, R.~R., {Brandt}, W.~N., {Gallagher}, S.~C.,
  {Hewett}, P.~C., \& {Schneider}, D.~P. 2010, \apj, 713, 220

\bibitem[{{Gibson}\noopsort{Gibsone} \& {Brandt}(2012)}]{2012ApJ...746...54G}
{Gibson}\noopsort{Gibsone}, R.~R., \& {Brandt}, W.~N. 2012, \apj, 746, 54

\bibitem[{{Green}\noopsort{Greena} \& {Mathur}(1996)}]{1996ApJ...462..637G}
{Green}\noopsort{Greena}, P.~J., \& {Mathur}, S. 1996, \apj, 462, 637

\bibitem[{{Green}\noopsort{Greenb} {et~al.}(2001){Green}\noopsort{Greenb},
  {Aldcroft}, {Mathur}, {Wilkes}, \& {Elvis}}]{2001ApJ...558..109G}
{Green}\noopsort{Greenb}, P.~J., {Aldcroft}, T.~L., {Mathur}, S., {Wilkes},
  B.~J., \& {Elvis}, M. 2001, \apj, 558, 109

\bibitem[{{Grupe} {et~al.}(2003){Grupe}, {Mathur}, \&
  {Elvis}}]{2003AJ....126.1159G}
{Grupe}, D., {Mathur}, S., \& {Elvis}, M. 2003, \aj, 126, 1159

\bibitem[{{G{\"u}ltekin} {et~al.}(2009){G{\"u}ltekin}, {Richstone}, {Gebhardt},
  {Lauer}, {Tremaine}, {Aller}, {Bender}, {Dressler}, {Faber}, {Filippenko},
  {Green}, {Ho}, {Kormendy}, {Magorrian}, {Pinkney}, \&
  {Siopis}}]{2009ApJ...698..198G}
{G{\"u}ltekin}, K., {et~al.} 2009, \apj, 698, 198

\bibitem[{{Hines} {et~al.}(2001){Hines}, {Schmidt}, {Gordon}, {Smith}, {Wills},
  {Allen}, \& {Sitko}}]{2001ApJ...563..512H}
{Hines}, D.~C., {Schmidt}, G.~D., {Gordon}, K.~D., {Smith}, P.~S., {Wills},
  B.~J., {Allen}, R.~G., \& {Sitko}, M.~L. 2001, \apj, 563, 512

\bibitem[{{Imanishi} \& {Terashima}(2004)}]{2004AJ....127..758I}
{Imanishi}, M., \& {Terashima}, Y. 2004, \aj, 127, 758

\bibitem[{{Just} {et~al.}(2007){Just}, {Brandt}, {Shemmer}, {Steffen},
  {Schneider}, {Chartas}, \& {Garmire}}]{2007ApJ...665.1004J}
{Just}, D.~W., {Brandt}, W.~N., {Shemmer}, O., {Steffen}, A.~T., {Schneider},
  D.~P., {Chartas}, G., \& {Garmire}, G.~P. 2007, \apj, 665, 1004

\bibitem[{{Kellermann} {et~al.}(1989){Kellermann}, {Sramek}, {Schmidt},
  {Shaffer}, \& {Green}}]{1989AJ.....98.1195K}
{Kellermann}, K.~I., {Sramek}, R., {Schmidt}, M., {Shaffer}, D.~B., \& {Green},
  R. 1989, \aj, 98, 1195

\bibitem[{{Kirsch} {et~al.}(2005){Kirsch}, {Briel}, {Burrows}, {Campana},
  {Cusumano}, {Ebisawa}, {Freyberg}, {Guainazzi}, {Haberl}, {Jahoda},
  {Kaastra}, {Kretschmar}, {Larsson}, {Lubi{\'n}ski}, {Mori}, {Plucinsky},
  {Pollock}, {Rothschild}, {Sembay}, {Wilms}, \&
  {Yamamoto}}]{2005SPIE.5898...22K}
{Kirsch}, M.~G., {et~al.} 2005, in Society of Photo-Optical Instrumentation
  Engineers (SPIE) Conference Series, Vol. 5898, Society of Photo-Optical
  Instrumentation Engineers (SPIE) Conference Series, ed. {O.~H.~W.~Siegmund},
  22--33

\bibitem[{{Laor}\noopsort{Laora} {et~al.}(1997){Laor}\noopsort{Laora}, {Fiore},
  {Elvis}, {Wilkes}, \& {McDowell}}]{1997ApJ...477...93L}
{Laor}\noopsort{Laora}, A., {Fiore}, F., {Elvis}, M., {Wilkes}, B.~J., \&
  {McDowell}, J.~C. 1997, \apj, 477, 93

\bibitem[{{Laor}\noopsort{Laorb} \& {Brandt}(2002)}]{2002ApJ...569..641L}
{Laor}\noopsort{Laorb}, A., \& {Brandt}, W.~N. 2002, \apj, 569, 641

\bibitem[{{Lawrence} {et~al.}(1997){Lawrence}, {Elvis}, {Wilkes}, {McHardy}, \&
  {Brandt}}]{1997MNRAS.285..879L}
{Lawrence}, A., {Elvis}, M., {Wilkes}, B.~J., {McHardy}, I., \& {Brandt}, N.
  1997, \mnras, 285, 879

\bibitem[{{Maloney} \& {Reynolds}(2000)}]{2000ApJ...545L..23M}
{Maloney}, P.~R., \& {Reynolds}, C.~S. 2000, \apjl, 545, L23

\bibitem[{{Manners} {et~al.}(2002){Manners}, {Almaini}, \&
  {Lawrence}}]{2002MNRAS.330..390M}
{Manners}, J., {Almaini}, O., \& {Lawrence}, A. 2002, \mnras, 330, 390

\bibitem[{{Mathur} {et~al.}(2000){Mathur}, {Green}, {Arav}, {Brotherton},
  {Crenshaw}, {deKool}, {Elvis}, {Goodrich}, {Hamann}, {Hines}, {Kashyap},
  {Korista}, {Peterson}, {Shields}, {Shlosman}, {van Breugel}, \&
  {Voit}}]{2000ApJ...533L..79M}
{Mathur}, S., {et~al.} 2000, \apjl, 533, L79

\bibitem[{{Miller} {et~al.}(2006){Miller}, {Brandt}, {Gallagher}, {Laor},
  {Wills}, {Garmire}, \& {Schneider}}]{2006ApJ...652..163M}
{Miller}, B.~P., {Brandt}, W.~N., {Gallagher}, S.~C., {Laor}, A., {Wills},
  B.~J., {Garmire}, G.~P., \& {Schneider}, D.~P. 2006, \apj, 652, 163

\bibitem[{{Mukai}(1993)}]{1993Legac...3...21M}
{Mukai}, K. 1993, Legacy, vol.~3, p.21--31, 3, 21

\bibitem[{{Murray}\noopsort{Murraya} {et~al.}(1995){Murray}\noopsort{Murraya},
  {Chiang}, {Grossman}, \& {Voit}}]{1995ApJ...451..498M}
{Murray}\noopsort{Murraya}, N., {Chiang}, J., {Grossman}, S.~A., \& {Voit},
  G.~M. 1995, \apj, 451, 498

\bibitem[{{Murray}\noopsort{Murrayb} \& {Chiang}(1997)}]{1997ApJ...474...91M}
{Murray}\noopsort{Murrayb}, N., \& {Chiang}, J. 1997, \apj, 474, 91

\bibitem[{{Nandra} {et~al.}(1997){Nandra}, {George}, {Mushotzky}, {Turner}, \&
  {Yaqoob}}]{1997ApJ...476...70N}
{Nandra}, K., {George}, I.~M., {Mushotzky}, R.~F., {Turner}, T.~J., \&
  {Yaqoob}, T. 1997, \apj, 476, 70

\bibitem[{{Nevalainen} {et~al.}(2010){Nevalainen}, {David}, \&
  {Guainazzi}}]{2010A&A...523A..22N}
{Nevalainen}, J., {David}, L., \& {Guainazzi}, M. 2010, \aap, 523, A22

\bibitem[{{Page} {et~al.}(2011){Page}, {Carrera}, {Stevens}, {Ebrero}, \&
  {Blustin}}]{2011MNRAS.416.2792P}
{Page}, M.~J., {Carrera}, F.~J., {Stevens}, J.~A., {Ebrero}, J., \& {Blustin},
  A.~J. 2011, \mnras, 416, 2792

\bibitem[{{Petric} {et~al.}(2006){Petric}, {Carilli}, {Bertoldi}, {Beelen},
  {Cox}, \& {Omont}}]{2006AJ....132.1307P}
{Petric}, A.~O., {Carilli}, C.~L., {Bertoldi}, F., {Beelen}, A., {Cox}, P., \&
  {Omont}, A. 2006, \aj, 132, 1307

\bibitem[{{Proga} {et~al.}(2000){Proga}, {Stone}, \&
  {Kallman}}]{2000ApJ...543..686P}
{Proga}, D., {Stone}, J.~M., \& {Kallman}, T.~R. 2000, \apj, 543, 686

\bibitem[{{Richards} {et~al.}(2011){Richards}, {Kruczek}, {Gallagher}, {Hall},
  {Hewett}, {Leighly}, {Deo}, {Kratzer}, \& {Shen}}]{2011AJ....141..167R}
{Richards}, G.~T., {et~al.} 2011, \aj, 141, 167

\bibitem[{{Rigopoulou} {et~al.}(1996){Rigopoulou}, {Lawrence}, \&
  {Rowan-Robinson}}]{1996MNRAS.278.1049R}
{Rigopoulou}, D., {Lawrence}, A., \& {Rowan-Robinson}, M. 1996, \mnras, 278,
  1049

\bibitem[{{Risaliti}\noopsort{Risalitia}
  {et~al.}(2002){Risaliti}\noopsort{Risalitia}, {Elvis}, \&
  {Nicastro}}]{2002ApJ...571..234R}
{Risaliti}\noopsort{Risalitia}, G., {Elvis}, M., \& {Nicastro}, F. 2002, \apj,
  571, 234

\bibitem[{{Risaliti}\noopsort{Risalitib}
  {et~al.}(2009){Risaliti}\noopsort{Risalitib}, {Salvati}, {Elvis}, {Fabbiano},
  {Baldi}, {Bianchi}, {Braito}, {Guainazzi}, {Matt}, {Miniutti}, {Reeves},
  {Soria}, \& {Zezas}}]{2009MNRAS.393L...1R}
{Risaliti}\noopsort{Risalitib}, G., {et~al.} 2009, \mnras, 393, L1

\bibitem[{{Ruiz} {et~al.}(2007){Ruiz}, {Carrera}, \&
  {Panessa}}]{2007A&A...471..775R}
{Ruiz}, A., {Carrera}, F.~J., \& {Panessa}, F. 2007, \aap, 471, 775

\bibitem[{{Saez} \& {Chartas}(2011)}]{2011ApJ...737...91S}
{Saez}, C., \& {Chartas}, G. 2011, \apj, 737, 91

\bibitem[{{Schartel}\noopsort{Shaa} {et~al.}(2005){Schartel}\noopsort{Shaa},
  {Rodr{\'{\i}}guez-Pascual}, {Santos-Lle{\'o}}, {Clavel}, {Guainazzi},
  {Jim{\'e}nez-Bail{\'o}n}, \& {Piconcelli}}]{2005A&A...433..455S}
{Schartel}\noopsort{Shaa}, N., {Rodr{\'{\i}}guez-Pascual}, P.~M.,
  {Santos-Lle{\'o}}, M., {Clavel}, J., {Guainazzi}, M.,
  {Jim{\'e}nez-Bail{\'o}n}, E., \& {Piconcelli}, E. 2005, \aap, 433, 455

\bibitem[{{Schartel}\noopsort{Shab} {et~al.}(2007){Schartel}\noopsort{Shab},
  {Rodr{\'{\i}}guez-Pascual}, {Santos-Lle{\'o}}, {Ballo}, {Clavel},
  {Guainazzi}, {Jim{\'e}nez-Bail{\'o}n}, \& {Piconcelli}}]{2007A&A...474..431S}
{Schartel}\noopsort{Shab}, N., {Rodr{\'{\i}}guez-Pascual}, P.~M.,
  {Santos-Lle{\'o}}, M., {Ballo}, L., {Clavel}, J., {Guainazzi}, M.,
  {Jim{\'e}nez-Bail{\'o}n}, E., \& {Piconcelli}, E. 2007, \aap, 474, 431

\bibitem[{{Schartel}\noopsort{Shac} {et~al.}(2010){Schartel}\noopsort{Shac},
  {Rodr{\'{\i}}guez-Pascual}, {Santos-Lle{\'o}}, {Jim{\'e}nez-Bail{\'o}n},
  {Ballo}, \& {Piconcelli}}]{2010A&A...512A..75S}
{Schartel}\noopsort{Shac}, N., {Rodr{\'{\i}}guez-Pascual}, P.~M.,
  {Santos-Lle{\'o}}, M., {Jim{\'e}nez-Bail{\'o}n}, E., {Ballo}, L., \&
  {Piconcelli}, E. 2010, \aap, 512, A75

\bibitem[{{Shemmer} {et~al.}(2005){Shemmer}, {Brandt}, {Gallagher}, {Vignali},
  {Boller}, {Chartas}, \& {Comastri}}]{2005AJ....130.2522S}
{Shemmer}, O., {Brandt}, W.~N., {Gallagher}, S.~C., {Vignali}, C., {Boller},
  T., {Chartas}, G., \& {Comastri}, A. 2005, \aj, 130, 2522

\bibitem[{{Sim} {et~al.}(2010){Sim}, {Proga}, {Miller}, {Long}, \&
  {Turner}}]{2010MNRAS.408.1396S}
{Sim}, S.~A., {Proga}, D., {Miller}, L., {Long}, K.~S., \& {Turner}, T.~J.
  2010, \mnras, 408, 1396

\bibitem[{{Snowden}(2002)}]{2002astro.ph..3311S}
{Snowden}, S.~L. 2002, ArXiv:astro-ph/0203311

\bibitem[{{Steffen} {et~al.}(2006){Steffen}, {Strateva}, {Brandt}, {Alexander},
  {Koekemoer}, {Lehmer}, {Schneider}, \& {Vignali}}]{2006AJ....131.2826S}
{Steffen}, A.~T., {Strateva}, I., {Brandt}, W.~N., {Alexander}, D.~M.,
  {Koekemoer}, A.~M., {Lehmer}, B.~D., {Schneider}, D.~P., \& {Vignali}, C.
  2006, \aj, 131, 2826

\bibitem[{{Tsujimoto} {et~al.}(2011){Tsujimoto}, {Guainazzi}, {Plucinsky},
  {Beardmore}, {Ishida}, {Natalucci}, {Posson-Brown}, {Read}, {Saxton}, \&
  {Shaposhnikov}}]{2011A&A...525A..25T}
{Tsujimoto}, M., {et~al.} 2011, \aap, 525, A25

\bibitem[{{Vanden Berk} {et~al.}(2001)}]{2001AJ....122..549V}
{Vanden Berk}, D.~E., {et~al.} 2001, \aj, 122, 549

\bibitem[{{Wang} {et~al.}(1996){Wang}, {Brinkmann}, \&
  {Bergeron}}]{1996A&A...309...81W}
{Wang}, T., {Brinkmann}, W., \& {Bergeron}, J. 1996, \aap, 309, 81

\bibitem[{{Weisskopf} {et~al.}(2010){Weisskopf}, {Guainazzi}, {Jahoda},
  {Shaposhnikov}, {O'Dell}, {Zavlin}, {Wilson-Hodge}, \&
  {Elsner}}]{2010ApJ...713..912W}
{Weisskopf}, M.~C., {Guainazzi}, M., {Jahoda}, K., {Shaposhnikov}, N.,
  {O'Dell}, S.~L., {Zavlin}, V.~E., {Wilson-Hodge}, C., \& {Elsner}, R.~F.
  2010, \apj, 713, 912

\bibitem[{{White} {et~al.}(1997){White}, {Becker}, {Helfand}, \&
  {Gregg}}]{1997ApJ...475..479W}
{White}, R.~L., {Becker}, R.~H., {Helfand}, D.~J., \& {Gregg}, M.~D. 1997,
  \apj, 475, 479

\bibitem[{{Wilkes} {et~al.}(1994){Wilkes}, {Tananbaum}, {Worrall}, {Avni},
  {Oey}, \& {Flanagan}}]{1994ApJS...92...53W}
{Wilkes}, B.~J., {Tananbaum}, H., {Worrall}, D.~M., {Avni}, Y., {Oey}, M.~S.,
  \& {Flanagan}, J. 1994, \apjs, 92, 53

\bibitem[{{Wu} {et~al.}(2010){Wu}, {Brandt}, {Comins}, {Gibson}, {Shemmer},
  {Garmire}, \& {Schneider}}]{2010ApJ...724..762W}
{Wu}, J., {Brandt}, W.~N., {Comins}, M.~L., {Gibson}, R.~R., {Shemmer}, O.,
  {Garmire}, G.~P., \& {Schneider}, D.~P. 2010, \apj, 724, 762

\end{thebibliography}

\providecommand{\noopsort}[1]{}

\end{document}